**Simulation-Guided Planning of a Target Trial Emulated Cluster Randomized Trial for Mass Small-Quantity Lipid Nutrient Supplementation Combined with Expanded Program on Immunization in Rural Niger**

**Authors:**
Shomoita Alam,[a] Nathaniel Dyrkton,[a] Susan Shepherd,[b] Ibrahim Sana,[b] Kevin Phelan,[b] and Jay JH Park[a,c]

**Affiliations:**
a. Core Clinical Sciences, Vancouver, BC, Canada
b. The Alliance for International Medical Action (ALIMA), Dakar, Senegal
c. Department of Health Research Methods, Evidence, and Impact, McMaster University, Hamilton, ON, Canada

**Corresponding author:**
Jay JH Park
Department of Health Research Methods, Evidence and Impact
McMaster University
1280 Main St W, Hamilton, ON L8S 4L8
+1 604-779-8240
parkj136@mcmaster.ca

## Abstract

**Background:** Target trial emulation (TTE) that applies trial design principles to improve the analysis of non-randomized studies is increasingly being used. Applications of TTE to emulate cluster randomized trials (RCTs) have been limited. This study explored how to integrate simulation-guided design into the TTE framework to inform planning of a non-randomized cluster trial.

**Methods:** We performed simulations to prospectively plan data collection of a non-randomized study emulating a village-level cluster RCT when cluster-randomization was infeasible. The planned study will assess the impact of mass distribution of nutritional supplements embedded within an existing immunization program to improve pentavalent vaccination rates among children 12-24 months old in Niger. The design included covariate-constrained random selection of villages for outcome ascertainment at follow-up. Simulations used baseline census data on pentavalent vaccination rates and cluster-level covariates to compare the type I error rate and power of four statistical methods: beta-regression; quasi-binomial regression; inverse probability of treatment weighting (IPTW); and naïve Wald test.

**Results:** Of the four analytic methods considered, only IPTW and beta-regression controlled the type I error rate at 0.05, but IPTW yielded poor statistical power. Beta-regression that showed adequate statistical power was chosen as our primary analysis.

**Conclusions:** Adopting simulation-guided design principles within TTE can enable robust planning of a group-level non-randomized study emulating a cluster RCT. Lessons from this study also apply to TTE planning of individually-RCTs.

**Keywords:** Target trial emulation; cluster randomized trial; simulation-guided design

**Funding:** This work was supported by the Gates Foundation, Seattle, WA USA.

## What is new?

### Key findings

- This study explored how to integrate simulation-guided design into the target trial emulation (TTE) framework to inform planning of a prospective, non-randomized cluster trial that aims to evaluate the impact of mass distribution of small-quantity lipid-based nutrient supplements in tandem with an expanded program on immunization on childhood vaccination coverage in rural Niger.

- By combining the TTE framework with simulation-guided design, we were able to identify, and then pre-specify in the trial statistical analysis plan, the optimal analytical method to estimate the effect of the intervention on vaccination coverage while controlling type I error and achieving target power.

### What this adds to what was known?

- This study demonstrates the value of simulation-guided design can reduce analytic bias in non-randomized TTE settings by clarifying the implications of assignment mechanism, covariate structures, and number of clusters affect estimator performance.

### What is the implication and what should change now?

- Future TTE studies can benefit from adopting simulation-guided design to improve estimator selection, pre-specification and overall validity of causal inferences.

- Incorporating simulation into TTE planning can strengthen alignment with trial design principles and enhance the robustness of analyses in settings where randomization is infeasible.

## Highlights

- A novel application of target trial emulation to a cluster design, beyond individual trials.
- Incorporating simulation-guided design can strengthen target trial emulation planning.
- Simulations help evaluate the performance of different estimators under specific design features.

## Introduction

Target trial emulation (TTE) is a framework that aims to improve the analysis of non-randomized data by applying clinical trial design principles.[1,2] TTE follows a two-step approach. First, a causal question of interest is specified, often in the form of a protocol of a hypothetical target trial of interest with specification of eligibility criteria, treatment strategies, assignment procedure, follow-up, outcomes, causal contrasts (i.e., causal estimands), and a statistical analysis plan.[2,3] These components of the target trial are emulated using non-randomized data. While the application of TTE to non-randomized studies has increased in recent years, the majority of applications have emulated individually randomized clinical trials (RCTs).[4] For public health and other interventions that are implemented at a group-level, individual-level randomization is frequently not feasible and can pose important risks for intervention contamination, leading to bias in effect estimation. In these instances, cluster RCTs are often used because they can improve feasibility of implementation, reduce the risk contamination between intervention groups, and, depending on the scale of implementation, can estimate population-level treatment effects that reflect real-world intervention delivery.

Given their non-randomized nature, TTE studies require measures to control for both measured and unmeasured confounding. To control for measured confounding, covariate-adjustment methods, such as propensity score matching and weighting, are commonly used.[1,5] For unmeasured confounding, the TTE framework recommends the use of negative control outcomes that are expected to not have any causal relationship with the intervention of interest.[6-8] Importantly, the data on confounders and negative control outcomes are often used in an iterative manner to decide on which analytical method and statistical estimator can best attenuate or remove bias.[6,9,10] These iterations during the planning stage may require multiple analyses be performed on the same dataset. While these steps are important for assessing and addressing bias and confounding in TTE studies, their use without detailed pre-specification contradicts the principles of trial design.

In trial design, it is recommended that details on confounding adjustments including specifications of covariates should carefully prespecified prior to any unblinding of trial data.[11] This is an important step to improve confidence in trial findings, especially in the primary analysis, to avoid ad-hoc changes after seeing the unblinded data.[12,13] In clinical trials, measures for blinding and control of information flow are also implemented to reduce potential operational biases, and ad-hoc changes are discouraged.[14] Speaking to their importance, together these principles comprise one of five bias domains assessed by the Risk of Bias 2.0, a widely recommended tool for assessing bias in RCTs.[15,16] In the Risk of Bias 2.0, failing to meet these criteria will designate a trial as being at high risk of bias. The absence of these practices in TTE, then, raises concern about bias mitigation and affords an opportunity for improvement by prespecifying all analyses.

A simulation-guided design is recommended to support rigorous trial planning.[17] We can balance the trade-offs between trial design options by exploring their performance across varying potential scenarios. For example, simulations can be used to compare different analytical methods during the planning stage to optimize the trial design. The utility of simulation-guided designs naturally extends to the planning of TTE by similarly allowing for comparisons of different analytical methods for non-randomized data under realistic assumptions without having to unblind the data. However, simulation-guided designs are rarely used in planning TTE studies. Indeed, it is entirely absent from recent reviews, descriptive papers, and guidance on the TTE framework.[1-4,6,18-21]

We aimed to use the TTE framework in tandem with statistical simulations to inform the design of a prospective, non-randomized intervention study called OptiMAx-Niger. OptiMAx-Niger is a multi-level clustered non-randomized trial that uses the TTE framework to evaluate the impact of mass distribution of small-quantity lipid-based nutrient supplements (SQ-LNS), embedded with an existing immunization program, on childhood vaccination coverage in rural remote villages within Mirriah, Niger. SQ-LNS is hypothesized to act as an incentive for caregivers to uptake childhood vaccines.

# Methods

In this section, we provide an overview of OptiMAx-Niger to situate our work. We discuss our target trial specification, target estimand and the TTE framework (detailed in Table 1) using the Transparent Reporting of Observational Studies Emulating a Target Trial (TARGET) guidelines[3]. Finally, we describe the simulation study we conducted to inform the data analysis plan for our TTE.

## *Overview of OptiMAx-Niger Study*

OptiMAx-Niger is a non-randomized, pre-post study that aims to determine the effectiveness of SQ-LNS coupled with delivery of vaccines within EPI to improve vaccine coverage in villages in Mirriah, Niger. The primary endpoint is the village-level proportion of children aged 12-24 months without a pentavalent vaccine dose (Penta0). Penta0 is used as a proxy measure for children who have received no vaccinations.[22]

The TTE framework was applied to OptiMAx-Niger to emulate a village-level cluster randomized trial. Village-level cluster randomization could not be performed without concerns of contamination due to the proximity of the villages. Randomization at the health area-level was also not possible due to concern about intervention contamination. Specifically, health center catchments, housed within health areas, are not defined by distance, and as a result some villages are not assigned to the closest health center.

A population-level baseline census was conducted between December 2024 to January 2025 across all villages in Mirriah, Niger. The census provided population-level measures of overall population size; number of children aged 6-59 months; pentavalent, measles and malaria vaccine coverage; and the distance to the nearest functional health center offering vaccinations. Population sizes for 5 villages were imputed by dividing the number of children aged 6-59 months by the average non-missing proportion of the number of children aged 6-59 months to the population size. The roll-out of SQ-LNS was planned for shortly after the completion of the baseline survey and upon receipt of ethics approval.

## *Target Estimand:*

In the hypothetical cluster trial, the primary estimand is a population-level treatment policy effect comparing villages receiving SQ-LNS+EPI versus EPI alone. The estimand is defined by: (i) population: all eligible children aged 12–24 months residing in villages meeting the inclusion criteria; (ii) treatment conditions: assignment to SQ-LNS+EPI versus EPI alone; (iii) endpoint: village-level proportion of children with no pentavalent vaccine dose 12 months after rollout; (iv) intercurrent events: all post-distribution events are considered part of the treatment policy strategy; (v) summary measure: the between-arm contrast in mean village-level Penta0 proportion.

In the emulation, this estimand corresponds to a population-average causal effect at the village level. Different analytical approaches target different estimands: IPTW targets a marginal contrast, while regression-based methods target conditional contrasts given village-level covariates. The goal of the simulations is not to estimate the causal effect itself, but to identify the analytic approach that best controls type I error and achieves acceptable power when used to estimate the pre-specified population-level estimand.

## ***Data Analysis & Simulation Study***

### *Analysis of Baseline Data*

We analyzed the baseline survey to parameterize the simulation study. A logistic regression model was fit to the baseline data with distance to the nearest medical center and village population as covariates. We use the estimates and 95% confidence intervals from this model as parameters for our data generating mechanism. The details are provided in Supplementary Materials 1.

### *Simulations*

We adopted a simulation-guided approach to select the best model to estimate the causal contrast of interest following the ADEMP (Aims, Data-generating mechanisms, Estimands, Methods, and Performance measures) scheme (simulation protocol provided in the Supplementary Materials 2).[24] We simulated 1,000,000 possible allocations of $n$ villages under a 1:1 randomization ratio, with the number of villages sampled from each health area proportional to the total number of villages in that area. For each allocation, covariate balance between treatment arms was assessed using SMDs for village-level covariates (total population, distance to the nearest health center, and baseline Penta0 rate), and only allocations with all SMDs $\leq 0.2$ were retained. Outcome data were then generated from a linear mixed-effects model parameterized using the ICC estimated from the baseline census and the WHO recommendation of 1/3 for planning vaccination surveys.[25] The base-case scenario was defined by a baseline vaccination rate of 0.20, regression coefficients set to the lower 95% confidence limits from the logistic regression analysis, and the empirically estimated ICC. In total, 360 distinct scenarios were investigated. Full details of the parameter values considered are provided in **Table 2**. The details of software used for computation are listed in Supplementary Materials 1.

### *Ethics*

Ethics approval for OptiMAx-Niger was obtained from the Comite National d'Ethique pour la Recherche en Sante (No33/2025/CNERS) in Niamey.

## Results

### *Baseline Census Survey Results*

The summary statistics of villages with at least five children aged 12-24 months in the baseline survey conducted in Mirriah, Niger are presented in **Supplementary Table S1**. Villages in the Eastern region (Group 1) were, on average, located farther from the nearest health center than those in the Western region (Group 2). Although the mean village population size was smaller in Group 1 (mean = 560.7, SD = 592.5) compared with Group 2 (mean = 1,017.3, SD = 901.2), the total number of eligible children was greater in Group 1 (5,153) than in Group 2 (4,376). The baseline Penta0 vaccination rate was slightly lower in Group 1 than in Group 2 (0.21 vs. 0.24, respectively; **Figure 1**).

Analysis of the baseline census using logistic regression to model baseline Penta0 rates yielded point estimate coefficients of -0.0001 and 0.0749 for village population and distance to the nearest health center respectively. The village level ICC was estimated to be 0.22. A full description of the simulation parameters is provided in **Table 2** and **Supplementary Table S2**.

### *Simulation Results*

In the base-case scenario, the expected type I error rate exceeded 0.05 for quasi-binomial regression and the naïve analysis (0.10 to 0.12, and 0.08 to 0.15, respectively; **Table 3**). In contrast, the beta regression maintained appropriate type I error control near 0.05. The IPTW approach was markedly conservative with type I error $< 0.01$. Similar patterns were observed across the other scenarios (**Supplementary Table S3**).

The target power was 80% at a one-sided type I error rate of 0.05. Under the base-case, both the naïve analysis and quasi-binomial regression achieved this target for detecting a 50% relative reduction in the Penta0 rate when at least 50 villages were included. With more than 50 villages, these approaches yielded empirical power exceeding 0.90 (see **Supplementary Figures S1 and S7**).[33] In contrast, the beta regression achieved adequate power to detect a 50% relative reduction only when at least 75 villages were selected, with power increasing as the number of villages to be selected increased (**Figure 2**). The IPTW approach exhibited consistently lower power than the other methods.

For a moderate effect size, defined as a 37.5% relative reduction in the Penta0 rate (identified by the ALIMA team as clinically meaningful), the base-case analysis using the beta regression model indicated that a sample size of 126 villages per arm would provide 80% power to detect this effect.

## Discussion

In this study, we used statistical simulations within the TTE framework to compare the performance of multiple estimators for a prospective, non-randomized cluster trial with covariate-constrained random selection of villages. Simulations were used to plan the emulation of this complex design. In the base-case, with a control event rate of 0.20, naïve and quasi-binomial analyses did not control the type I error rate at 0.05 (ranges: 0.08-0.15, and 0.10- 0.12, respectively). The inflated type I error rate for the naïve model was expected, as covariate adjustment of variables used in covariate-constrained randomization is recommended to maintain the type I error rate.[34,35] Conversely, our simulations showed that propensity-score method in our case study was overly conservative, with power under 0.50, regardless of the number of villages sampled. Beta regression provided adequate type I error control and sufficient power to detect a clinically important effect size with a feasible number of villages in OptiMAx-Niger.

This study demonstrates the utility of simulation-guided design for planning TTE analyses. Through simulations, we assessed the performance of covariate-constrained random selection aimed at emulating a covariate-constrained randomization procedure and evaluated type I error control and power across multiple scenarios. This was essential for identifying an optimal estimator for the clinical and implementation context of OptiMAx-Niger. Although IPTW is common in TTE,[21] our simulations showed that it would be overly conservative for OptiMAx-Niger, yielding excessive type I error control and inadequate power.

To our knowledge, this is the first emulation of a cluster randomized trial planned using simulation-guided design principles. This may reflect the fact that the TTE framework is often applied to retrospective observational studies. Here, however, we demonstrate its value for careful design of non-randomized experimental studies with prospective data collection.

By pairing the TTE framework with a simulation-guided design, we assessed whether our pre-specified covariates were sufficient to control measured confounding, a central concern in non-randomized studies. Recent work has extended simulation methods to evaluate the impact of unmeasured confounding in non-randomized studies.[36] Together these approaches could strengthen TTE-recommended practices for detecting unmeasured confounding and residual bias, including the use of negative outcome controls.[1,3,18] Although negative outcome controls are well established in the life sciences and epidemiology,[7-10] their implementation within TTE, and the actions to be taken if bias is detected, are often not clearly defined or pre-specified. For example, a recent review of TTE-based observational studies found that only 12% of the studies reviewed had an available protocol.[21] While current reporting guidelines include bias assessment,[3] the required level of detail is insufficient to meet trial principles of pre-specification, allowing selective reporting.[37] Adopting simulation-guided design in TTE studies would facilitate pre-specification, reduce bias and improve alignment with RCT design principles, even in complex settings, such as cluster trials.

The findings from this study are strengthened by robust methodology and interdisciplinary collaboration. Our simulations were grounded in a baseline census, providing confidence in our estimated relationships between key covariates and the primary outcome. In addition, the study was designed in collaboration with local and international stakeholders with deep expertise in vaccination interventions and contextual knowledge of rural and remote regions of Niger. Together, these elements enabled the development of a robust, fit-for-purpose design.

This work has limitations. Due to practical constraints in data collection, we were limited in the number of covariates that could be balanced through covariate-constrained random selection, leaving risk of unmeasured confounding. Because treatment assignment followed geographic health-area implementation, residual confounding is plausible even after adjustment for village population, distance to health centre, and baseline Penta0 rate. Structural differences across areas, such as health-system infrastructure, caregiver mobility and socioeconomic conditions, may not be fully captured in the baseline census. If access to services is systematically poorer in the control areas, confounding could exaggerate the apparent benefit of SQ-LNS; conversely, greater vaccination outreach capacity in intervention areas could attenuate observed effects. As detailed in the statistical analysis plan (see Supplementary Materials 3), we will conduct quantitative bias analysis to assess the impact of unmeasured confounding, and the validity of the conditional exchangeability assumption.[38,39] To evaluate the robustness to violations of this assumption, we will calculate the E-value, defined as the minimum strength of association an unmeasured confounder would need with both treatment (SQ-LNS vs. EPI) and outcome (number of children aged 12-24 months without a pentavalent vaccine dose) to fully explain the observed effect, conditional on measured covariates.[40,41]

We compared statistical estimators targeting different estimands (e.g., marginal vs conditional treatment effects). The simulations aimed to design a TTE study that maximized statistical power while controlling the type I error rate 0.05. Although the IPTW estimator targets a marginal effect, it was markedly conservative in our setting, with inflated standard errors and reduced power. This behavior is consistent with prior work showing that sandwich variance estimators used in IPTW analyses can overestimate uncertainty from propensity score estimation, yielding confidence intervals with greater-than-nominal coverage, sometimes substantially so, particularly when treatment is non-randomized and the propensity score depends on covariates predictive of treatment but not outcome.[42,43] In contrast, beta regression achieved nominal type I error and superior operating characteristics across all scenarios and was therefore selected as the prespecified primary analysis. These findings underscore the need for further methodological research on weighting-based estimators in clustered, non-randomized settings, where variance inflation and weight instability may be amplified.

### ***Future Directions***

This study has important implications for future research on public health interventions and patient populations. Although cluster-level randomization is often more feasible than individual-level randomization, there remains settings in which randomization is not possible due to ethical or practical considerations. In OptiMAx-Niger, for example, the proximity of health areas and villages within made randomization infeasible because of substantial contamination risk. By using covariate-constrained random selection, we achieved balance between study arms and identified an estimator with optimal operating characteristics. These methods can inform future policy studies seeking reliable estimates of intervention impact when randomization, even at the cluster-level, is not possible.

Consideration of study operating characteristics *a priori* is imperative for ethical research, regardless of whether prospective or retrospective data are used. By enabling optimization of operating characteristics, simulation-guided design helps ensure that only the number of participants necessary to answer the primary research question is used. When simulations indicate that an emulated trial cannot be adequately powered, researchers can avoid analyses that would yield uninformative results.

Future studies within the TTE framework can build on this work by implementing simulation-guided design to fully pre-specify analyses and assess operating characteristics. Broader adoption of these methods in TTE will reduce bias from subjective data assessment and selective reporting and strengthen confidence in results from non-randomized studies. Given the importance of robust study design, simulation-guided design should be promoted as an evidence-based method to enhance TTE research.

## Conclusions

This study extends simulation-guided design to TTE of a cluster randomized trial evaluating the impact of nutritional supplement distribution on vaccine coverage in remote and rural regions of Niger. Grounded in a comprehensive baseline census, interdisciplinary expertise and simulation, we designed a robust, non-randomized study with optimal type I error control and power and that aligned with operational constraints. These methods are feasible and provide valuable insights for research in non-randomized settings.

## Declaration of Interests

The authors have no conflicts of interest to declare.

## Author Contributions

SA contributed to the conceptualization and design of the study, drafted the initial manuscript and provided revisions. ND contributed to the conceptualization and design of the study, drafted the initial manuscript and provided revisions. IS contributed to data curation, investigation and project administration. KP and SS contributed to the conceptualization and design of this work and provided oversight and revisions to the manuscript. JJHP contributed to the conceptualization and design of the study, conducted analyses, contributed to manuscript drafting and provided revisions.

## Data Sharing Statement

Requests for data can be submitted to the corresponding author and will be considered by the author group in conjunction with ALIMA’s leadership in Niger.

# Tables

**Table 1: Reporting the hypothetical cluster trial and target trial emulation using the Transparent Reporting of Observational Studies Emulating a Target Trial (TARGET) guidelines**

| Component | Hypothetical cluster trial | Target trial emulation |
| --- | --- | --- |
| **Eligibility Criteria** | There would be village- and participant-level eligibility criteria for the target trial. The same eligibility criteria would be applied for the follow-up survey.<br><br>*Inclusion:* At the village-level, we would consider any villages located in the 10 health areas of interest with at least 5 children aged 12-24 months as of the baseline survey. Within each eligible village, any child aged 12-24 months of age with oral informed-consent from caregivers residing in the catchment settlements would be eligible.<br><br>*Exclusion:* Villages with fewer than five children aged 12-24 months would be excluded as they represent observations with high uncertainty in the proportion of children without a pentavalent vaccine dose. | Same as hypothetical trial. |
| **Treatment strategies** | There would be two arms in the target trial.<br><br>*Control arm:* The control arm would be the standard EPI.<br><br>*Intervention:* The intervention arm would be SQ-LNS delivered within the standard EPI. | Same as hypothetical trial. |
| **Assignment procedures** | We would randomly select villages from the participating health areas and then use covariate-constrained randomization to assign selected villages to receive either the intervention or control.<br><br>The standardized mean difference (SMDs) between the village-level covariates (distance to nearest health center, population, and baseline Penta0 rates) of the treatment and control arms would be restricted to ≤0.2. | As randomization was not possible, treatment was determined by practical considerations with Group 1 health areas (Zermou, Guéza Mahaman, Kissambana, Hamdara, Angoual Malan) receiving the control and Group 2 health areas (Danéki, Droum, Incharoua, Kabda, and Magaria Toukour) receiving the intervention. |
| **Follow-up period** | A cross-sectional survey of villages would be conducted 12 months after the start of SQ-LNS distribution. | A cross-sectional survey of villages will be conducted 12 months after the start of SQ-LNS distribution. Surveying all villages at follow-up was assessed to be infeasible, so we will apply covariate-constrained random selection based on baseline census data to determine which villages to sample at follow-up and to prevent imbalance between the study arms.<br><br>As in the targe trial, the SMDs between the village-level covariates (distance to nearest health center, population, and baseline Penta0 rates) of the treatment and control arms will be restricted to ≤ 0.2. |
| **Outcomes** | Our primary endpoint would be Penta0 measured at the 12-month post follow-up survey, which is the proportion of children aged 12-24 months with no pentavalent vaccination.<br><br>Vaccination status would be confirmed by card or caregiver-report. | Same as hypothetical trial. |
| **Causal contrasts** | We would be interested in the treatment policy effects of the SQ-LNS distribution defined as the difference in pentavalent vaccine coverage between villages that receive SQ-LNS coupled with EPI and those that receive only EPI, measured as a risk difference (RD).<br><br>Any events occurring after the distribution of SQ-LNS that could affect the primary endpoints would be considered part of the intervention. Let $\pi_a$ be the probability of a child in village a receiving vaccine where the village $a = 1$ receives SQ-LNS and $a = 0$ does not. Our summary effect measure would be defined as: RD = ($\pi_1$ - $\pi_0$). | The causal contrast of interest in our target trial emulation is the observational analogue of our specified target trial contrast of interest: treatment policy effects of the SQ-LNS distribution measured as the difference in pentavalent vaccine coverage between villages that receive SQ-LNS coupled with EPI and those that receive only EPI measured as an RR.<br><br>As in the target trial, any events occurring after the distribution of SQ-LNS that may affect the primary endpoint will be considered part of the intervention. |
| **Assumptions** | Due to the design of our target trial and our outcome ascertainment methods, we determined that no assumptions about loss to follow-up would be needed. | We assumed conditional exchangeability. In other words, it was assumed that villages were exchangeable between treatment groups conditioned on the baseline vaccination rate, the total population and the distance to the nearest health center. |
| **Data analysis plan** | We would adopt a beta regression approach and fit a beta generalized linear model with mean-precision parameterization. As randomization was covariate constrained, the analysis would account for constraining covariates as is recommended to avoid inflating type I error.[a,b] We would test the hypothesis: $H_0$: $\beta = 0$ vs. $H_1$: $\beta < 0$. A Wald test would be performed with the nominal estimate and cluster-robust standard error from the beta regression fit. The clustering would done by village. If $H_0$ were rejected, we would conclude that the distribution of the nutritional supplement has a significant effect on Penta0 rates. | Our data analysis plan will be determined based on a simulation study to compare the operating characteristics of four competing models to operationalize our causal estimand: beta regression; quasi-binomial regression; inverse probability of treatment weighting (IPTW); and a naïve Wald test without any covariate adjustments. Further details are provided in Supplementary Materials 1. |

[a] Ciolino JD, Schauer JM, Bonner LB. Covariate-Constrained Randomization. *JAMA Intern Med*. Jun 30 2025; doi: 10.1001/jamainternmed.2025.2566
[b] Moulton LH. Covariate-based constrained randomization of group-randomized trials. *Clin Trials*. 2004;1(3):297-305. doi:10.1191/1740774504cn024oa

**Table 2: Summary of factors used in the data-generating mechanism**

| Factor | Value | Justification |
|---|---|---|
| **Conditional relative reduction in Penta0 rate** | $\delta_r$ = 0, 0.15, 0.25, 0.375, 0.5 | Effect sizes of interest |
| **Follow-up Penta0 rate in the control arm** | $\pi_0$ = 0.15, 0.2, 0.25, 0.3 | Expert opinion and baseline survey |
| **Number of villages sampled per arm** | $n$ = 50, 75, 80, 100, 110, 126 | The minimum sample size was chosen based on preliminary sample size calculations of a cluster RCT. The maximum sample size was constrained by the number of villages in the smaller arm of each study |
| **Main effect of village population** | $\beta_1$ = -0.00010860, -0.00015608, -0.00006112 | Logistic regression of baseline Penta0 rates<br>Lower 95% CI, point estimate, Upper 95% CI |
| **Main effect of distance of village to nearest health center** | $\beta_2$ = 0.074920, 0.061783, 0.088057 | Logistic regression of baseline Penta0 rates<br>Lower 95% CI, point estimate, Upper 95% CI |
| **Intra-cluster correlation (ICC)** | $ICC = 0.22, \frac{1}{3}$ | Approximated ICC calculated based on the baseline survey according and WHO recommendation for planning vaccination surveys |

**Table 3: Estimates and 95% confidence intervals of type I error rate of different methods under the base-case (Penta0 rate of 0.20 in the control arm, coefficient set 1, ICC of 0.22)**

| Villages per arm | Quasi-binomial | Beta | IPTW | Naive |
|---|---|---|---|---|
| 50 | 0.109 (0.103, 0.115) | 0.050 (0.046, 0.055) | 0.009 (0.007, 0.011) | 0.080 (0.074, 0.085) |
| 75 | 0.102 (0.096, 0.108) | 0.041 (0.037, 0.045) | 0.008 (0.006, 0.009) | 0.102 (0.096, 0.108) |
| 80 | 0.112 (0.106, 0.118) | 0.046 (0.042, 0.050) | 0.007 (0.006, 0.009) | 0.107 (0.101, 0.113) |
| 100 | 0.104 (0.098, 0.110) | 0.042 (0.038, 0.046) | 0.006 (0.005, 0.008) | 0.119 (0.112, 0.125) |
| 110 | 0.115 (0.109, 0.122) | 0.042 (0.038, 0.046) | 0.007 (0.005, 0.008) | 0.132 (0.125, 0.138) |
| 126 | 0.109 (0.103, 0.115) | 0.044 (0.040, 0.048) | 0.004 (0.003, 0.005) | 0.151 (0.144, 0.158) |

## Figures

**Figure 1: Village-level baseline distribution of Penta0 rates obtained from census data by arm**

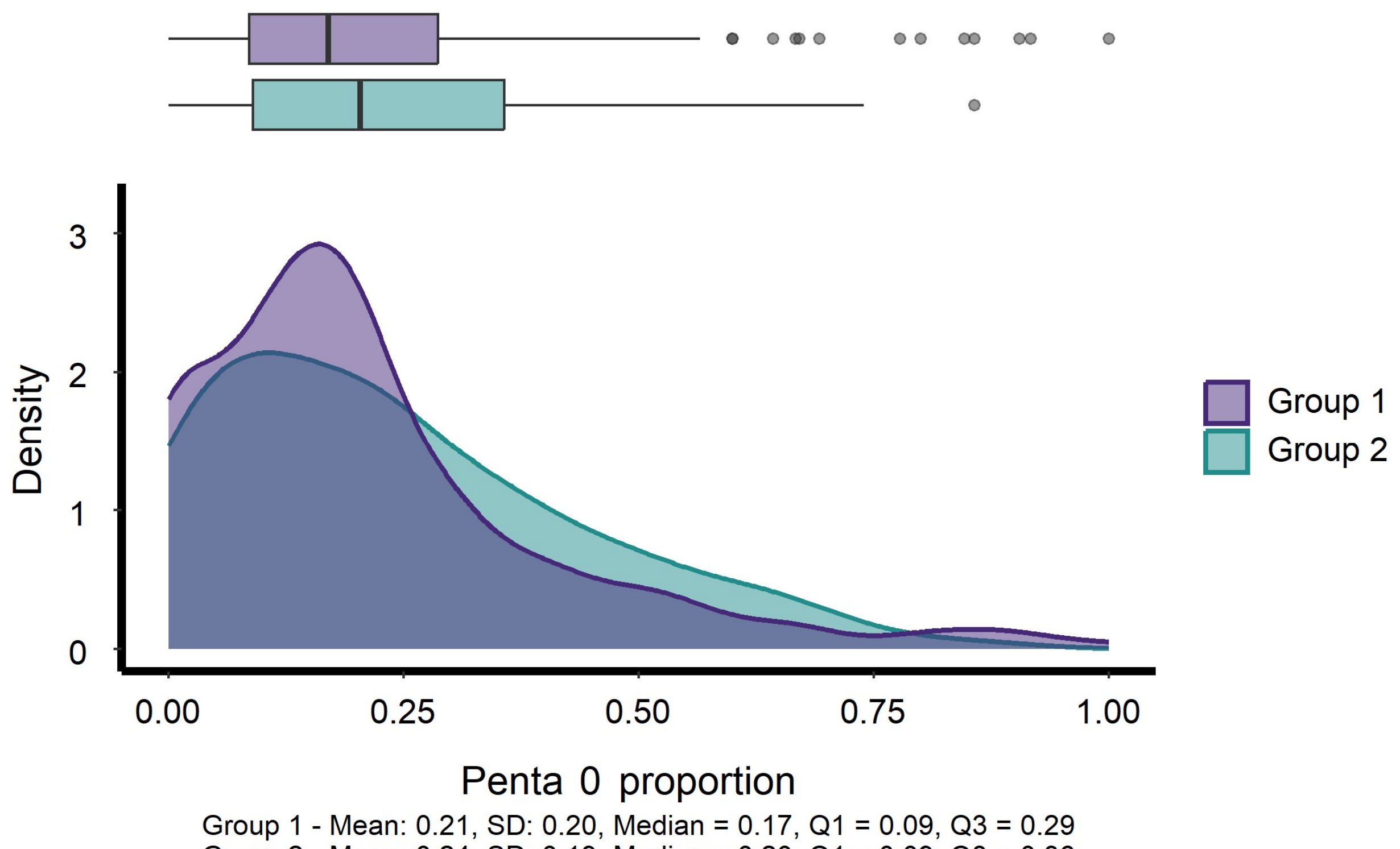

**Figure 2: Power of beta regression at different sample sizes and relative reductions in Penta0 rate under the base-case (Penta0 rate of 0.20 in the control arm, coefficient set 1, ICC of 0.22)**

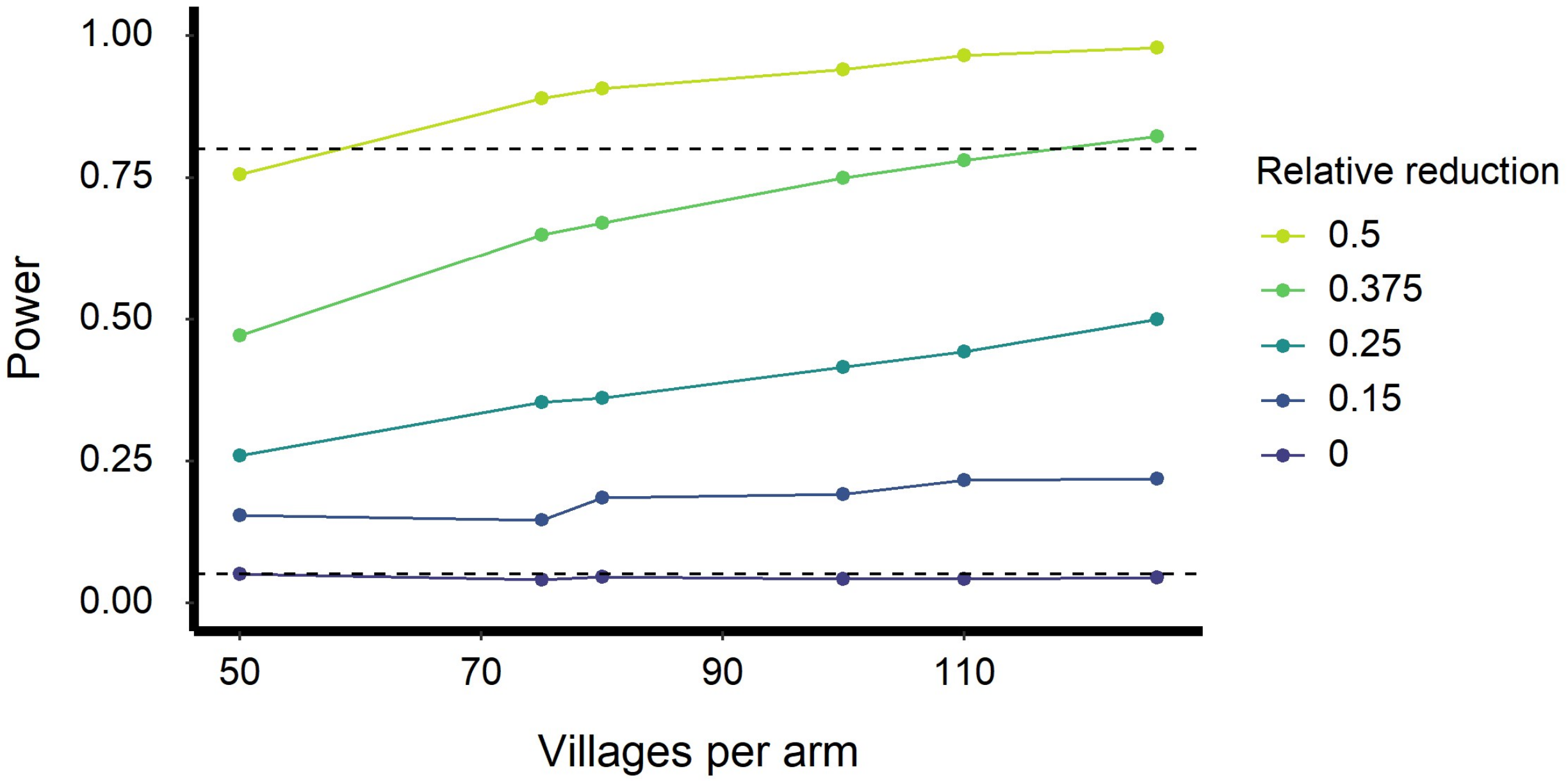

## Supplementary Materials:

Supplementary to *“Simulation-Guided Planning of a Target Trial Emulation for a Cluster Randomized Trial for Mass Small-Quantity Lipid Nutrient Supplementation Combined with Expanded Program on Immunization in Rural Niger”*

# 1. Supplementary Methods

## *Details of Baseline Census Data Analysis for Simulation Parameters including the Intracluster Correlation Coefficient (ICC) Calculations*

Let $Y_j^0$ be the total number of children aged 12-24 months with Penta0 and let $m_j^0$ be the total number of children aged 12-24 months in village $j$. Let $p_j$ be the population at the baseline survey, and $d_j$ be the distance to the nearest health center, then the logistic regression model is

$$Y_j^0 \sim Bin\left(m_j^0, {\pi'}_j\right),$$

$$\log\frac{{\pi'}_j}{1-{\pi'}_j} = \eta_0 + \eta_1 p_j + \eta_2 d_j.$$

The coefficient estimates from the fitted models were used to inform the coefficient values for our simulation study. We also used the baseline census-like data to obtain an estimate of the ICC at the village level using the following model

$$Y_j^0 \mid \nu_j \sim Bin\left(m_j^0, {\pi'}_j\right),$$

$$\log\frac{{\pi'}_j}{1-{\pi'}_j} = \eta_0 + \nu_j + \eta_1 p_j + \eta_2 d_j, \nu_j \sim N(0, \tau^2),$$

where $p_j$ and $d_j$ are the population and distance of village $j$ respectively. To avoid convergence issues, we standardize $p_j$ into $\tilde{p}_j$

$$\tilde{p}_j = \frac{p_j - \overline{p}}{sd(p)},$$

where $\overline{p}$ and $sd(p)$ were the mean and standard deviation of total village population. Lastly, to estimate the ICC for binary outcomes,(1) we estimate it using the variance of the village level random effect

$$ICC = \frac{\hat{\tau}^2}{\frac{\pi^2}{3} + \hat{\tau}^2},$$

where $\pi^2/3$ is the residual variance of the logistic distribution.

### *Simulation Details*

In each simulation replicate, a random pair of health area and village with an average village-level SMD not more than 0.2 was chosen. The number of vaccinated children aged 12-24 months that would be available for the follow-up survey were generated according to the health area assignment of this pair. Subsequently, the data were generated via a mixed-effects logistic regression. Let $Y_j^t$ be the number of children aged 12-24 months with Penta0 in village $j$ at time $t$, where $t = 0$ is the baseline census, and $t = 1$ is the follow-up survey. Then the $Y_j^1$ was generated as follows:

$$Y_j^1 \sim Bin\left(m_j^1, \pi_j^1\right),$$

$$\log\frac{\pi_j^1}{1-\pi_j^1} = \beta_0 + \alpha_j + \log\frac{Y_j^0}{m_j^0 - Y_j^0} + \beta a_j + \beta_1 p_j + \beta_2 d_j, \alpha_j \sim N(0, \tau^2).$$

Within each repetition we varied the relative decrease of Penta0 ($\delta_a$= 0,0.15,0.25,0.375,0.5); the follow-up Penta0 rate in the control arm ($\pi_0$ = 0.15,0.2,0.25,0.3); the number of villages sampled per arm ($n$); the main effect of village population ($\beta_1$ = -0.00010860, -0.00015608, -0.00006112; the main effect of distance to nearest health centre ($\beta_2$ = 0.074920,0.061783,0.088057); and the ICC (0.22, 1/3). The three values for the coefficients are the lower 95% confidence interval, point estimate, and upper 95% confidence interval values, respectively, for each parameter. The first ICC value was estimated from the baseline survey data and the second was the conservative recommendation for the World Health Organization's vaccination survey planning guidance.(2) The follow-up Penta0 rates in the control arm were based on expert opinion.

We also further defined coefficient set $i$ as $i^{th}$ value of $\beta_1$ and $\beta_2$. The combination of all these coefficient sets corresponded to 360 scenarios per repetition. We defined the base-case simulation scenario as: a Penta0 rate of 0.20 in the control arm; village population coefficient of -0.00010860; distance to nearest health centre coefficient of -0.074920; and an ICC of 0.22 while varying the number of villages sampled per arm.

### *Monte Carlo Error*

Any simulation study involves uncertainty in our estimates in using a finite number of replications. When estimating the Type 1 error and Power, we utilized 10,000 and 1,000 repetitions respectively. To quantify the error in our Monte Carlo simulation we use results for the Monte Carlo standard errors.(3) Let $n_{rep}$ be the number of repetitions in our simulation and we assume the null hypothesis is true (assuming an absolute increase of vaccine coverage =0), then our point estimate for the type 1 error is:

$$\widehat{\text{Type 1}}\text{ error} = \frac{1}{n_{rep}} \sum_{k=1}^{n_{rep}} I\,(p_k \leq \alpha)$$

with the Monte Carlo Standard Error of

$$\widehat{MCSE}_{\widehat{\text{Type 1}}\text{ error}} = \sqrt{\frac{\widehat{\text{Type 1}}\text{ error}\left(1 - \widehat{\text{Type 1}}\text{ error}\right)}{n_{rep}}}$$

Thus, using the central limit theorem, we then arrive at

$$\widehat{\text{Type 1}}\text{ error} \sim \text{Normal}\left(\text{Type 1 error}, \frac{\widehat{\text{Type 1}}\text{ error}\left(1 - \widehat{\text{Type 1}}\text{ error}\right)}{n_{rep}}\right)$$

and we can then construct the corresponding 95% Monte Carlo confidence intervals. We may also construct the same confidence interval for power by setting the absolute increase in vaccination rate $> 0$ and using the same formulas.

### *Operationalization of Causal Estimands*

Let $Y_j^t$ be the number of children aged 12-24 months in village $j = 1, \ldots, n$, without a pentavalent vaccine dose at the baseline survey if $t = 0$ and at the follow-up survey if $t = 1$, $m_j^t$ be the number of children aged 12-24 months at the baseline survey if $t = 0$ and at the follow-up survey if $t = 1$, $a_j$ be the 0-1 indicator for whether SQ-LNS is distributed after the baseline survey, $p_j$ be the population at the baseline survey, and $d_j$ be the distance to the nearest health center. We standardized $p_j$ prior to analysis to avoid convergence issues. We also added an interaction term between $p_j$ and $d_j$ to express our uncertainty about the data generating mechanism.

For our first model, we adopted a beta regression approach.(4, 5) The model followed a beta generalized linear model with mean-precision parametrization.

$$\frac{Y_j^1}{m_j^1} \sim Beta(\pi_j, \varphi),$$

$$Var\left(\frac{Y_j^1}{m_j^1}\right) = \frac{\pi_j(1-\pi_j)}{1+\varphi},$$

$$\log\frac{\pi_j}{1-\pi_j} = \beta_0 + \beta a_j + \beta_1\frac{Y_j^0}{m_j^0} + \beta_2 p_j + \beta_3 d_j + \beta_4 p_j d_j.$$

We transform $\frac{Y_j^1}{m_j^1}$ such that it lies strictly within the interval (0,1). (6) Our second model used quasi-binomial regression. (7)

$$\mathbb{E}Y_j^1 = m_j^1\pi_j(1-\pi_j)$$

$$Var(Y_j^1) = \varphi m_j^1\pi_j(1-\pi_j)$$

$$\log\frac{\pi_j}{1-\pi_j} = \beta_0 + \beta a_j + \beta_1\frac{Y_j^0}{m_j^0} + \beta_2 p_j + \beta_3 d_j + \beta_4 p_j d_j$$

In both the quasi-binomial and beta regression, we will test the hypothesis $H_0$: $\beta = 0$ vs. $H_1$: $\beta < 0$.

Our third model will use propensity score weights.(8) Let $Y_{\cdot,a}^1$ be the potential outcome to denote the number of children aged 12-24 months without a pentavalent vaccine dose at the follow-up survey in an arbitrary village for treatment status $a = 0,1$, and let $m^1$ be the total number of children aged 12-24 months in the same village at the follow-up survey. We define our probability of vaccination as

$$\pi_{\cdot,a} = \mathbb{E}\frac{Y_{\cdot,a}^1}{m^1},$$

and our contrast as $\delta = \pi_{\cdot,1} - \pi_{\cdot,0}$.We then tested the hypothesis $H_0$: $\delta = 0$ vs. $H_1$: $\delta < 0$. We estimated $\delta$ with $\hat{\delta} = \hat{\pi}_{\cdot,1} - \hat{\pi}_{\cdot,0}$, and in turn, we estimate $\pi_{\cdot,a}$ with

$$\hat{\pi}_{\cdot,a} = \sum_{i=1}^{n}\widehat{w}_j\frac{Y_j^1 I(a_j = a)}{m_j^1},$$

where

$$\widehat{w}_j = \frac{1}{\widehat{P}\left(A_j = 1 \mid Y_j^0, m_j^0, p_j, d_j\right)A_j + \widehat{P}\left(A_j = 0 \mid Y_j^0, m_j^0, p_j, d_j\right)\left(1 - A_j\right)},$$

with the conditional probabilities estimated by fitting a logistic regression model as

$$\log\frac{P\left(A_j = 1 \mid Y_j^0, m_j^0, p_j, d_j\right)}{P\left(A_j = 0 \mid Y_j^0, m_j^0, p_j, d_j\right)} = \gamma_0 + \gamma_1\frac{Y_j^0}{m_j^0} + \gamma_2 p_j + \gamma_3 d_j + \gamma_4 p_j d_j + \gamma_5 p_j d_j.$$

The standard error of the treatment effect estimator was calculated using the robust heteroscedastic-consistent variance estimator assuming known weights.(9) Our simulations suggested that a critical value of -1.811911, corresponding to a one-sided significance level of 0.035, should be used to control the Type I error rate at 0.05. This critical value was used in all analyses.

### *Computation*

All simulations were conducted in R version 4.4.0. Data management was performed using the ***dplyr*** and ***tidyr*** packages, and graphical outputs were generated with ***ggplot2***.(10) (11) (12) Mixed-effects logistic regression models were fit using ***lme4***, beta regression models were fit using ***betareg***, and robust variance estimators for the inverse probability of treatment weighting analyses were obtained using ***sandwich***. (13) (4, 5) (14) Simulations were executed using the Google Cloud computing platform.

### *Sensitivity Analysis*

The association conveyed by the E-value is on the risk ratio scale and pertains to the relationships between the unmeasured confounder, the treatment and the outcome only. More specifically, suppose $\hat{\beta}$ is the estimate of the conditional odds ratio of not having received a dose of the pentavalent vaccine as measured in the follow-up survey. The approximate E-value for an odds ratio (15) is

$$E = \sqrt{b} + \sqrt{\sqrt{b}\left(\sqrt{b} - 1\right)},$$

where

$$b = \max\left(\hat{\beta}, \frac{1}{\hat{\beta}}\right).$$

E-values can be used to assess the minimum strength of association that an unmeasured confounder would need to have between the treatment groups for the observed treatment effect to become null (16). E-values can be interpreted in the context of the observed treatment effect. Broadly, E-values larger than the treatment effect can indicate that effect estimates are robust against unmeasured confounding, while smaller E-values indicate otherwise.

For our sensitivity analyses, we will evaluate the magnitude of the E-values on the risk ratio scale instead of the odds ratio scale for easier interpretability.

With $R_{CT}$ being the risk ratio of the confounder between treatment groups and $R_{CO}$ being the risk ratio of the outcome between levels of the confounder, we calculate the following bias-adjusted effect estimate (16):

$$\hat{\beta}_{adj} = \hat{\beta} \cdot \left(\frac{R_{CT} + R_{CO} - 1}{R_{CT} R_{CO}}\right)^2.$$

# 2. Supplementary Tables

***Supplementary Table S1. Descriptive statistics of villages in Mirriah, Niger by arm***

| Characteristic | Group 1 - East, EPI (control)<br>N of villages = 224 | Group 2 - West, SQ-LNS (intervention)<br>N of villages = 126 |
|---|---|---|
| Mean (SD) of village distance to nearest health center[1] | 6.6 (3.8) | 5.8 (3.3) |
| Mean (SD) of total village population size[2] | 560.7 (592.5) | 1,017.3 (901.2) |
| Mean (SD) of number of children aged 12-24 months in village[3] | 23.0 (20.5) | 34.7 (30.9) |
| Total number of children aged 12-24 months in group[4] | 5,153 | 4,376 |
| Total number (proportion) of children aged 12-24 months with Penta0 in group[5] | 1,144 (0.22) | 967 (0.22) |
| Mean (SD) of village Penta0 rate[6] | 0.21 (0.20) | 0.24 (0.19) |
| Median (Q1, Q3) of village Penta0 rate | 0.17 (0.09, 0.29) | 0.20 (0.09, 0.36) |

[1]Distance of a village to the nearest health center, derived from GPS information during data collection

[2]Total population of a village, derived from census data

[3]Number of children aged 12-24 months in a village as recorded in baseline survey

[4]Sum of number of children aged 12-24 months across all villages in each group

[5]Sum of number of children aged 12-24 months with Penta0 vaccination across all villages in each group. The proportion is the number of children aged 12-24 months with Penta0 vaccination divided by the number of children aged 12-24 months in total in each group

[6]Number of children aged 12-24 months with Penta0 vaccination in a village divided by number of children aged 12-24 months in the village, as recorded in baseline survey

***Supplementary Table S2. Coefficient sets for the $\beta_1$ and $\beta_2$ parameters used in the data-generating mechanism.***

| Coefficient set | Village population | Distance to nearest health center |
|---|---|---|
| 1 | -0.00010860 | 0.074920 |
| 2 | -0.00015608 | 0.061783 |
| 3 | -0.00006112 | 0.088057 |

***Supplementary Table S3. Estimates and 95% confidence intervals of type I error rate of different methods under varying scenarios, excluding the base case, based on control event rate, coefficient values, and intracluster correlation coefficient (ICC)***

| Control Penta0 rate: 0.15; Coefficient set: 1; ICC: 0.22 | | | | |
|---|---|---|---|---|
| **Villages per arm** | **Quasi-binomial** | **Beta** | **IPTW** | **Naive** |
| 50 | 0.103 (0.097, 0.109) | 0.056 (0.052, 0.061) | 0.010 (0.009, 0.012) | 0.103 (0.097, 0.109) |
| 75 | 0.099 (0.093, 0.105) | 0.052 (0.047, 0.056) | 0.008 (0.006, 0.010) | 0.134 (0.127, 0.140) |
| 80 | 0.106 (0.100, 0.112) | 0.050 (0.045, 0.054) | 0.009 (0.007, 0.011) | 0.141 (0.134, 0.147) |
| 100 | 0.107 (0.101, 0.113) | 0.052 (0.048, 0.057) | 0.008 (0.007, 0.010) | 0.168 (0.160, 0.175) |
| 110 | 0.102 (0.096, 0.108) | 0.051 (0.047, 0.055) | 0.006 (0.004, 0.008) | 0.186 (0.179, 0.194) |
| 126 | 0.106 (0.100, 0.112) | 0.051 (0.047, 0.055) | 0.004 (0.003, 0.006) | 0.211 (0.203, 0.219) |

| Control Penta0 rate: 0.15; Coefficient set: 1; ICC: 1/3 | | | | |
|---|---|---|---|---|
| **Villages per arm** | **Quasi-binomial** | **Beta** | **IPTW** | **Naive** |
| 50 | 0.100 (0.094, 0.106) | 0.044 (0.040, 0.048) | 0.014 (0.012, 0.016) | 0.086 (0.081, 0.092) |
| 75 | 0.099 (0.093, 0.105) | 0.040 (0.036, 0.044) | 0.010 (0.009, 0.012) | 0.104 (0.098, 0.110) |
| 80 | 0.099 (0.093, 0.105) | 0.038 (0.034, 0.042) | 0.011 (0.009, 0.013) | 0.106 (0.100, 0.112) |
| 100 | 0.103 (0.097, 0.109) | 0.037 (0.033, 0.040) | 0.010 (0.008, 0.012) | 0.129 (0.122, 0.135) |
| 110 | 0.101 (0.095, 0.106) | 0.042 (0.038, 0.046) | 0.008 (0.006, 0.009) | 0.141 (0.135, 0.148) |
| 126 | 0.098 (0.092, 0.104) | 0.041 (0.037, 0.045) | 0.007 (0.005, 0.009) | 0.157 (0.150, 0.164) |

| Control Penta0 rate: 0.15; Coefficient set: 2; ICC: 0.22 | | | | |
|---|---|---|---|---|
| **Villages per arm** | **Quasi-binomial** | **Beta** | **IPTW** | **Naive** |
| 50 | 0.102 (0.096, 0.107) | 0.053 (0.049, 0.058) | 0.011 (0.009, 0.013) | 0.112 (0.106, 0.118) |
| 75 | 0.105 (0.099, 0.112) | 0.054 (0.049, 0.058) | 0.007 (0.006, 0.009) | 0.154 (0.147, 0.161) |
| 80 | 0.111 (0.104, 0.117) | 0.056 (0.052, 0.061) | 0.009 (0.007, 0.011) | 0.161 (0.154, 0.168) |
| 100 | 0.106 (0.100, 0.112) | 0.054 (0.049, 0.058) | 0.009 (0.008, 0.011) | 0.190 (0.182, 0.198) |

| Control Penta0 rate: 0.15; Coefficient set: 1; ICC: 0.22 | | | | |
|---|---|---|---|---|
| **Villages per arm** | **Quasi-binomial** | **Beta** | **IPTW** | **Naive** |
| 110 | 0.106 (0.100, 0.113) | 0.052 (0.047, 0.056) | 0.006 (0.005, 0.008) | 0.208 (0.200, 0.216) |
| 126 | 0.105 (0.099, 0.111) | 0.051 (0.047, 0.055) | 0.005 (0.004, 0.007) | 0.233 (0.224, 0.241) |

| Control Penta0 rate: 0.15; Coefficient set: 2; ICC: 1/3 | | | | |
|---|---|---|---|---|
| **Villages per arm** | **Quasi-binomial** | **Beta** | **IPTW** | **Naive** |
| 50 | 0.105 (0.099, 0.111) | 0.043 (0.039, 0.047) | 0.016 (0.013, 0.018) | 0.086 (0.081, 0.092) |
| 75 | 0.104 (0.098, 0.110) | 0.041 (0.037, 0.045) | 0.012 (0.010, 0.015) | 0.117 (0.110, 0.123) |
| 80 | 0.107 (0.101, 0.113) | 0.042 (0.039, 0.046) | 0.012 (0.010, 0.015) | 0.120 (0.113, 0.126) |
| 100 | 0.113 (0.106, 0.119) | 0.042 (0.038, 0.046) | 0.010 (0.008, 0.012) | 0.146 (0.139, 0.153) |
| 110 | 0.107 (0.101, 0.114) | 0.042 (0.038, 0.046) | 0.008 (0.006, 0.010) | 0.155 (0.148, 0.162) |
| 126 | 0.105 (0.099, 0.111) | 0.043 (0.039, 0.047) | 0.006 (0.005, 0.008) | 0.172 (0.165, 0.180) |

| Control Penta0 rate: 0.15; Coefficient set: 3; ICC: 0.22 | | | | |
|---|---|---|---|---|
| **Villages per arm** | **Quasi-binomial** | **Beta** | **IPTW** | **Naive** |
| 50 | 0.102 (0.096, 0.108) | 0.052 (0.048, 0.057) | 0.010 (0.008, 0.012) | 0.090 (0.085, 0.096) |
| 75 | 0.106 (0.100, 0.112) | 0.054 (0.049, 0.058) | 0.011 (0.009, 0.013) | 0.131 (0.125, 0.138) |
| 80 | 0.112 (0.106, 0.118) | 0.053 (0.049, 0.057) | 0.009 (0.008, 0.011) | 0.136 (0.129, 0.142) |
| 100 | 0.108 (0.102, 0.114) | 0.053 (0.049, 0.057) | 0.008 (0.006, 0.010) | 0.154 (0.147, 0.161) |
| 110 | 0.106 (0.100, 0.112) | 0.054 (0.050, 0.058) | 0.008 (0.006, 0.009) | 0.176 (0.169, 0.184) |
| 126 | 0.107 (0.101, 0.113) | 0.053 (0.048, 0.057) | 0.006 (0.004, 0.007) | 0.194 (0.186, 0.202) |

| Control Penta0 rate: 0.15; Coefficient set: 3; ICC: 1/3 | | | | |
|---|---|---|---|---|
| **Villages per arm** | **Quasi-binomial** | **Beta** | **IPTW** | **Naive** |

| Control Penta0 rate: 0.15; Coefficient set: 1; ICC: 0.22 | | | | |
|---|---|---|---|---|
| **Villages per arm** | **Quasi-binomial** | **Beta** | **IPTW** | **Naive** |
| 50 | 0.101 (0.095, 0.106) | 0.044 (0.040, 0.048) | 0.011 (0.009, 0.014) | 0.077 (0.071, 0.082) |
| 75 | 0.103 (0.098, 0.109) | 0.043 (0.039, 0.047) | 0.012 (0.010, 0.014) | 0.105 (0.099, 0.111) |
| 80 | 0.103 (0.098, 0.109) | 0.045 (0.041, 0.049) | 0.011 (0.009, 0.014) | 0.107 (0.101, 0.113) |
| 100 | 0.110 (0.104, 0.116) | 0.042 (0.038, 0.045) | 0.011 (0.009, 0.013) | 0.125 (0.118, 0.131) |
| 110 | 0.110 (0.103, 0.116) | 0.042 (0.038, 0.046) | 0.009 (0.007, 0.011) | 0.135 (0.128, 0.141) |
| 126 | 0.112 (0.106, 0.119) | 0.046 (0.042, 0.050) | 0.008 (0.006, 0.009) | 0.156 (0.149, 0.164) |

| Control Penta0 rate: 0.20; Coefficient set: 1; ICC: 1/3 | | | | |
|---|---|---|---|---|
| **Villages per arm** | **Quasi-binomial** | **Beta** | **IPTW** | **Naive** |
| 50 | 0.100 (0.094, 0.105) | 0.038 (0.035, 0.042) | 0.011 (0.009, 0.013) | 0.068 (0.064, 0.073) |
| 75 | 0.105 (0.099, 0.111) | 0.035 (0.031, 0.039) | 0.011 (0.009, 0.013) | 0.083 (0.077, 0.088) |
| 80 | 0.108 (0.102, 0.114) | 0.034 (0.031, 0.038) | 0.009 (0.008, 0.011) | 0.082 (0.077, 0.087) |
| 100 | 0.112 (0.106, 0.118) | 0.031 (0.028, 0.034) | 0.009 (0.007, 0.011) | 0.097 (0.091, 0.103) |
| 110 | 0.103 (0.097, 0.109) | 0.030 (0.027, 0.034) | 0.007 (0.005, 0.009) | 0.101 (0.095, 0.107) |
| 126 | 0.110 (0.104, 0.117) | 0.036 (0.032, 0.040) | 0.008 (0.006, 0.010) | 0.110 (0.104, 0.116) |

| Control Penta0 rate: 0.20; Coefficient set: 2; ICC: 0.22 | | | | |
|---|---|---|---|---|
| **Villages per arm** | **Quasi-binomial** | **Beta** | **IPTW** | **Naive** |
| 50 | 0.102 (0.096, 0.108) | 0.044 (0.040, 0.049) | 0.009 (0.007, 0.010) | 0.085 (0.079, 0.090) |
| 75 | 0.111 (0.105, 0.118) | 0.045 (0.041, 0.049) | 0.008 (0.006, 0.010) | 0.103 (0.097, 0.109) |
| 80 | 0.106 (0.100, 0.112) | 0.045 (0.041, 0.050) | 0.008 (0.006, 0.010) | 0.107 (0.101, 0.113) |
| 100 | 0.110 (0.104, 0.117) | 0.042 (0.038, 0.046) | 0.006 (0.005, 0.008) | 0.130 (0.123, 0.136) |

| Control Penta0 rate: 0.15; Coefficient set: 1; ICC: 0.22 | | | | |
|---|---|---|---|---|
| **Villages per arm** | **Quasi-binomial** | **Beta** | **IPTW** | **Naive** |
| 110 | 0.106 (0.100, 0.112) | 0.041 (0.037, 0.045) | 0.004 (0.002, 0.005) | 0.131 (0.124, 0.138) |
| 126 | 0.116 (0.110, 0.122) | 0.043 (0.039, 0.047) | 0.005 (0.004, 0.006) | 0.153 (0.146, 0.160) |
| **Control Penta0 rate: 0.20; Coefficient set: 2; ICC: 1/3** | | | | |
| **Villages per arm** | **Quasi-binomial** | **Beta** | **IPTW** | **Naive** |
| 50 | 0.103 (0.097, 0.109) | 0.035 (0.031, 0.038) | 0.012 (0.010, 0.014) | 0.072 (0.067, 0.077) |
| 75 | 0.099 (0.093, 0.105) | 0.031 (0.028, 0.035) | 0.009 (0.007, 0.011) | 0.080 (0.075, 0.085) |
| 80 | 0.108 (0.102, 0.114) | 0.034 (0.030, 0.037) | 0.011 (0.009, 0.013) | 0.089 (0.084, 0.095) |
| 100 | 0.108 (0.102, 0.114) | 0.034 (0.030, 0.037) | 0.009 (0.007, 0.010) | 0.102 (0.096, 0.108) |
| 110 | 0.108 (0.102, 0.114) | 0.033 (0.029, 0.037) | 0.009 (0.007, 0.010) | 0.106 (0.100, 0.113) |
| 126 | 0.111 (0.105, 0.117) | 0.033 (0.029, 0.036) | 0.006 (0.004, 0.007) | 0.117 (0.111, 0.123) |
| **Control Penta0 rate: 0.20; Coefficient set: 3; ICC: 0.22** | | | | |
| **Villages per arm** | **Quasi-binomial** | **Beta** | **IPTW** | **Naive** |
| 50 | 0.104 (0.098, 0.110) | 0.048 (0.044, 0.052) | 0.009 (0.007, 0.011) | 0.077 (0.072, 0.083) |
| 75 | 0.109 (0.103, 0.115) | 0.046 (0.042, 0.050) | 0.010 (0.008, 0.012) | 0.095 (0.090, 0.101) |
| 80 | 0.110 (0.104, 0.116) | 0.046 (0.042, 0.050) | 0.009 (0.007, 0.010) | 0.096 (0.090, 0.101) |
| 100 | 0.109 (0.103, 0.115) | 0.042 (0.039, 0.046) | 0.006 (0.005, 0.008) | 0.110 (0.104, 0.116) |
| 110 | 0.112 (0.106, 0.118) | 0.043 (0.039, 0.047) | 0.006 (0.004, 0.007) | 0.120 (0.114, 0.127) |
| 126 | 0.110 (0.104, 0.116) | 0.041 (0.037, 0.045) | 0.005 (0.004, 0.007) | 0.136 (0.129, 0.143) |
| **Control Penta0 rate: 0.20; Coefficient set: 3; ICC: 1/3** | | | | |
| **Villages per arm** | **Quasi-binomial** | **Beta** | **IPTW** | **Naive** |

| Control Penta0 rate: 0.15; Coefficient set: 1; ICC: 0.22 | | | | |
|---|---|---|---|---|
| **Villages per arm** | **Quasi-binomial** | **Beta** | **IPTW** | **Naive** |
| 50 | 0.102 (0.096, 0.108) | 0.035 (0.032, 0.039) | 0.011 (0.009, 0.013) | 0.064 (0.059, 0.069) |
| 75 | 0.107 (0.101, 0.113) | 0.038 (0.034, 0.042) | 0.012 (0.010, 0.014) | 0.080 (0.075, 0.086) |
| 80 | 0.107 (0.101, 0.113) | 0.037 (0.033, 0.041) | 0.010 (0.008, 0.012) | 0.078 (0.073, 0.084) |
| 100 | 0.116 (0.109, 0.122) | 0.036 (0.032, 0.039) | 0.010 (0.008, 0.012) | 0.094 (0.089, 0.100) |
| 110 | 0.112 (0.106, 0.118) | 0.034 (0.031, 0.038) | 0.007 (0.006, 0.009) | 0.095 (0.089, 0.101) |
| 126 | 0.115 (0.109, 0.121) | 0.035 (0.031, 0.039) | 0.007 (0.005, 0.009) | 0.114 (0.107, 0.120) |

| Control Penta0 rate: 0.25; Coefficient set: 1; ICC: 0.22 | | | | |
|---|---|---|---|---|
| **Villages per arm** | **Quasi-binomial** | **Beta** | **IPTW** | **Naive** |
| 50 | 0.105 (0.099, 0.111) | 0.040 (0.036, 0.044) | 0.008 (0.006, 0.009) | 0.063 (0.058, 0.068) |
| 75 | 0.106 (0.100, 0.112) | 0.036 (0.033, 0.040) | 0.006 (0.005, 0.008) | 0.073 (0.068, 0.078) |
| 80 | 0.113 (0.107, 0.120) | 0.042 (0.039, 0.046) | 0.010 (0.008, 0.012) | 0.078 (0.073, 0.083) |
| 100 | 0.111 (0.105, 0.117) | 0.035 (0.031, 0.039) | 0.006 (0.005, 0.008) | 0.084 (0.078, 0.089) |
| 110 | 0.119 (0.112, 0.125) | 0.036 (0.032, 0.039) | 0.008 (0.006, 0.010) | 0.092 (0.086, 0.098) |
| 126 | 0.116 (0.110, 0.122) | 0.033 (0.030, 0.037) | 0.003 (0.002, 0.004) | 0.096 (0.090, 0.102) |

| Control Penta0 rate: 0.25; Coefficient set: 1; ICC: 1/3 | | | | |
|---|---|---|---|---|
| **Villages per arm** | **Quasi-binomial** | **Beta** | **IPTW** | **Naive** |
| 50 | 0.102 (0.096, 0.108) | 0.033 (0.029, 0.036) | 0.010 (0.008, 0.012) | 0.056 (0.051, 0.060) |
| 75 | 0.109 (0.103, 0.115) | 0.030 (0.027, 0.034) | 0.011 (0.009, 0.013) | 0.067 (0.062, 0.072) |
| 80 | 0.112 (0.106, 0.118) | 0.032 (0.029, 0.035) | 0.012 (0.010, 0.014) | 0.065 (0.060, 0.070) |
| 100 | 0.114 (0.108, 0.121) | 0.029 (0.026, 0.032) | 0.010 (0.008, 0.012) | 0.073 (0.068, 0.078) |

| Control Penta0 rate: 0.15; Coefficient set: 1; ICC: 0.22 | | | | |
|---|---|---|---|---|
| **Villages per arm** | **Quasi-binomial** | **Beta** | **IPTW** | **Naive** |
| 110 | 0.112 (0.106, 0.118) | 0.032 (0.029, 0.035) | 0.009 (0.007, 0.011) | 0.082 (0.077, 0.087) |
| 126 | 0.116 (0.109, 0.122) | 0.030 (0.027, 0.033) | 0.006 (0.005, 0.008) | 0.089 (0.083, 0.095) |
| **Control Penta0 rate: 0.25; Coefficient set: 2; ICC: 0.22** | | | | |
| **Villages per arm** | **Quasi-binomial** | **Beta** | **IPTW** | **Naive** |
| 50 | 0.100 (0.094, 0.106) | 0.040 (0.036, 0.044) | 0.006 (0.005, 0.008) | 0.058 (0.054, 0.063) |
| 75 | 0.110 (0.104, 0.116) | 0.038 (0.034, 0.041) | 0.007 (0.005, 0.009) | 0.082 (0.076, 0.087) |
| 80 | 0.113 (0.107, 0.120) | 0.040 (0.036, 0.044) | 0.007 (0.005, 0.009) | 0.081 (0.076, 0.087) |
| 100 | 0.110 (0.104, 0.117) | 0.036 (0.033, 0.040) | 0.005 (0.004, 0.007) | 0.093 (0.087, 0.099) |
| 110 | 0.118 (0.111, 0.124) | 0.037 (0.034, 0.041) | 0.006 (0.005, 0.008) | 0.105 (0.099, 0.111) |
| 126 | 0.121 (0.115, 0.128) | 0.037 (0.034, 0.041) | 0.005 (0.004, 0.006) | 0.111 (0.105, 0.117) |
| **Control Penta0 rate: 0.25; Coefficient set: 2; ICC: 1/3** | | | | |
| **Villages per arm** | **Quasi-binomial** | **Beta** | **IPTW** | **Naive** |
| 50 | 0.101 (0.095, 0.107) | 0.034 (0.031, 0.038) | 0.014 (0.012, 0.017) | 0.063 (0.058, 0.068) |
| 75 | 0.106 (0.100, 0.112) | 0.033 (0.030, 0.037) | 0.010 (0.009, 0.012) | 0.077 (0.072, 0.083) |
| 80 | 0.110 (0.104, 0.116) | 0.032 (0.029, 0.035) | 0.012 (0.010, 0.014) | 0.077 (0.072, 0.082) |
| 100 | 0.113 (0.107, 0.119) | 0.032 (0.028, 0.035) | 0.010 (0.008, 0.012) | 0.086 (0.081, 0.091) |
| 110 | 0.113 (0.107, 0.119) | 0.030 (0.026, 0.033) | 0.006 (0.005, 0.008) | 0.088 (0.082, 0.093) |
| 126 | 0.120 (0.114, 0.127) | 0.031 (0.027, 0.034) | 0.007 (0.005, 0.008) | 0.099 (0.093, 0.105) |
| **Control Penta0 rate: 0.25; Coefficient set: 3; ICC: 0.22** | | | | |
| **Villages per arm** | **Quasi-binomial** | **Beta** | **IPTW** | **Naive** |

| Control Penta0 rate: 0.15; Coefficient set: 1; ICC: 0.22 | | | | |
|---|---|---|---|---|
| **Villages per arm** | **Quasi-binomial** | **Beta** | **IPTW** | **Naive** |
| 50 | 0.103 (0.097, 0.109) | 0.036 (0.032, 0.039) | 0.005 (0.003, 0.006) | 0.050 (0.045, 0.054) |
| 75 | 0.107 (0.101, 0.113) | 0.036 (0.032, 0.039) | 0.006 (0.005, 0.008) | 0.056 (0.052, 0.061) |
| 80 | 0.112 (0.106, 0.118) | 0.033 (0.029, 0.036) | 0.008 (0.006, 0.009) | 0.055 (0.051, 0.059) |
| 100 | 0.119 (0.112, 0.125) | 0.034 (0.031, 0.038) | 0.007 (0.005, 0.008) | 0.064 (0.059, 0.069) |
| 110 | 0.114 (0.108, 0.121) | 0.034 (0.030, 0.037) | 0.006 (0.004, 0.007) | 0.068 (0.064, 0.073) |
| 126 | 0.117 (0.110, 0.123) | 0.030 (0.027, 0.034) | 0.005 (0.004, 0.007) | 0.076 (0.071, 0.081) |
| **Control Penta0 rate: 0.25; Coefficient set: 3; ICC: 1/3** | | | | |
| **Villages per arm** | **Quasi-binomial** | **Beta** | **IPTW** | **Naive** |
| 50 | 0.113 (0.107, 0.120) | 0.033 (0.030, 0.037) | 0.011 (0.009, 0.013) | 0.054 (0.050, 0.058) |
| 75 | 0.110 (0.104, 0.116) | 0.031 (0.028, 0.034) | 0.013 (0.011, 0.015) | 0.055 (0.050, 0.059) |
| 80 | 0.111 (0.105, 0.117) | 0.030 (0.026, 0.033) | 0.011 (0.009, 0.013) | 0.059 (0.054, 0.064) |
| 100 | 0.116 (0.110, 0.123) | 0.030 (0.027, 0.034) | 0.011 (0.009, 0.013) | 0.066 (0.061, 0.071) |
| 110 | 0.119 (0.112, 0.125) | 0.033 (0.029, 0.036) | 0.008 (0.007, 0.010) | 0.072 (0.067, 0.077) |
| 126 | 0.122 (0.115, 0.128) | 0.030 (0.027, 0.034) | 0.007 (0.006, 0.009) | 0.075 (0.070, 0.080) |
| **Control Penta0 rate: 0.30; Coefficient set: 1; ICC: 0.22** | | | | |
| **Villages per arm** | **Quasi-binomial** | **Beta** | **IPTW** | **Naive** |
| 50 | 0.112 (0.105, 0.118) | 0.037 (0.033, 0.040) | 0.008 (0.006, 0.009) | 0.043 (0.039, 0.047) |
| 75 | 0.116 (0.110, 0.122) | 0.033 (0.029, 0.037) | 0.007 (0.006, 0.009) | 0.050 (0.046, 0.054) |
| 80 | 0.121 (0.115, 0.127) | 0.035 (0.031, 0.038) | 0.009 (0.007, 0.010) | 0.053 (0.048, 0.057) |
| 100 | 0.123 (0.116, 0.129) | 0.032 (0.029, 0.036) | 0.008 (0.007, 0.010) | 0.056 (0.051, 0.060) |

| Control Penta0 rate: 0.15; Coefficient set: 1; ICC: 0.22 | | | | |
|---|---|---|---|---|
| **Villages per arm** | **Quasi-binomial** | **Beta** | **IPTW** | **Naive** |
| 110 | 0.129 (0.122, 0.135) | 0.033 (0.029, 0.036) | 0.007 (0.005, 0.008) | 0.060 (0.055, 0.064) |
| 126 | 0.133 (0.126, 0.140) | 0.033 (0.030, 0.037) | 0.006 (0.004, 0.008) | 0.063 (0.058, 0.068) |
| **Control Penta0 rate: 0.30; Coefficient set: 1; ICC: 1/3** | | | | |
| **Villages per arm** | **Quasi-binomial** | **Beta** | **IPTW** | **Naive** |
| 50 | 0.104 (0.098, 0.110) | 0.030 (0.026, 0.033) | 0.010 (0.008, 0.012) | 0.042 (0.038, 0.046) |
| 75 | 0.108 (0.102, 0.115) | 0.028 (0.025, 0.032) | 0.011 (0.009, 0.013) | 0.048 (0.044, 0.052) |
| 80 | 0.122 (0.116, 0.129) | 0.029 (0.026, 0.032) | 0.010 (0.008, 0.012) | 0.052 (0.048, 0.056) |
| 100 | 0.121 (0.114, 0.127) | 0.027 (0.024, 0.030) | 0.010 (0.008, 0.012) | 0.052 (0.048, 0.057) |
| 110 | 0.127 (0.120, 0.134) | 0.024 (0.021, 0.027) | 0.008 (0.006, 0.010) | 0.056 (0.052, 0.061) |
| 126 | 0.132 (0.126, 0.139) | 0.026 (0.023, 0.029) | 0.008 (0.006, 0.009) | 0.064 (0.059, 0.068) |
| **Control Penta0 rate: 0.30; Coefficient set: 2; ICC: 0.22** | | | | |
| **Villages per arm** | **Quasi-binomial** | **Beta** | **IPTW** | **Naive** |
| 50 | 0.104 (0.098, 0.110) | 0.035 (0.031, 0.038) | 0.006 (0.004, 0.008) | 0.043 (0.039, 0.047) |
| 75 | 0.116 (0.110, 0.122) | 0.034 (0.031, 0.038) | 0.008 (0.006, 0.009) | 0.053 (0.048, 0.057) |
| 80 | 0.122 (0.115, 0.128) | 0.037 (0.034, 0.041) | 0.006 (0.005, 0.008) | 0.056 (0.052, 0.061) |
| 100 | 0.117 (0.111, 0.123) | 0.034 (0.030, 0.037) | 0.008 (0.006, 0.010) | 0.062 (0.057, 0.067) |
| 110 | 0.128 (0.122, 0.135) | 0.035 (0.031, 0.038) | 0.005 (0.004, 0.007) | 0.064 (0.059, 0.069) |
| 126 | 0.129 (0.123, 0.136) | 0.031 (0.027, 0.034) | 0.005 (0.003, 0.006) | 0.068 (0.064, 0.073) |
| **Control Penta0 rate: 0.30; Coefficient set: 2; ICC: 1/3** | | | | |
| **Villages per arm** | **Quasi-binomial** | **Beta** | **IPTW** | **Naive** |

| Control Penta0 rate: 0.15; Coefficient set: 1; ICC: 0.22 | | | | |
|---|---|---|---|---|
| **Villages per arm** | **Quasi-binomial** | **Beta** | **IPTW** | **Naive** |
| 50 | 0.103 (0.097, 0.109) | 0.032 (0.028, 0.035) | 0.012 (0.010, 0.014) | 0.050 (0.046, 0.055) |
| 75 | 0.109 (0.103, 0.116) | 0.028 (0.025, 0.031) | 0.010 (0.008, 0.012) | 0.053 (0.048, 0.057) |
| 80 | 0.118 (0.111, 0.124) | 0.029 (0.026, 0.033) | 0.010 (0.008, 0.012) | 0.054 (0.050, 0.059) |
| 100 | 0.118 (0.112, 0.125) | 0.025 (0.022, 0.028) | 0.009 (0.007, 0.011) | 0.057 (0.053, 0.062) |
| 110 | 0.121 (0.114, 0.127) | 0.026 (0.023, 0.030) | 0.008 (0.006, 0.010) | 0.058 (0.053, 0.062) |
| 126 | 0.126 (0.120, 0.133) | 0.023 (0.020, 0.026) | 0.007 (0.005, 0.008) | 0.065 (0.060, 0.070) |
| **Control Penta0 rate: 0.30; Coefficient set: 3; ICC: 0.22** | | | | |
| **Villages per arm** | **Quasi-binomial** | **Beta** | **IPTW** | **Naive** |
| 50 | 0.108 (0.102, 0.114) | 0.036 (0.032, 0.040) | 0.007 (0.006, 0.009) | 0.040 (0.036, 0.044) |
| 75 | 0.117 (0.111, 0.124) | 0.035 (0.032, 0.039) | 0.008 (0.006, 0.010) | 0.050 (0.045, 0.054) |
| 80 | 0.120 (0.114, 0.127) | 0.033 (0.029, 0.037) | 0.007 (0.006, 0.009) | 0.044 (0.040, 0.048) |
| 100 | 0.133 (0.126, 0.140) | 0.034 (0.030, 0.038) | 0.007 (0.006, 0.009) | 0.053 (0.049, 0.058) |
| 110 | 0.128 (0.121, 0.135) | 0.031 (0.027, 0.034) | 0.008 (0.006, 0.009) | 0.052 (0.048, 0.056) |
| 126 | 0.133 (0.126, 0.139) | 0.034 (0.030, 0.037) | 0.006 (0.005, 0.008) | 0.056 (0.051, 0.060) |
| **Control Penta0 rate: 0.30; Coefficient set: 3; ICC: 1/3** | | | | |
| **Villages per arm** | **Quasi-binomial** | **Beta** | **IPTW** | **Naive** |
| 50 | 0.105 (0.099, 0.112) | 0.032 (0.028, 0.035) | 0.010 (0.008, 0.012) | 0.042 (0.038, 0.046) |
| 75 | 0.110 (0.104, 0.116) | 0.028 (0.025, 0.032) | 0.012 (0.010, 0.014) | 0.045 (0.041, 0.049) |
| 80 | 0.115 (0.109, 0.122) | 0.027 (0.024, 0.030) | 0.010 (0.008, 0.012) | 0.045 (0.041, 0.049) |
| 100 | 0.123 (0.117, 0.130) | 0.026 (0.023, 0.029) | 0.011 (0.009, 0.013) | 0.052 (0.048, 0.056) |

| Control Penta0 rate: 0.15; Coefficient set: 1; ICC: 0.22 | | | | |
|---|---|---|---|---|
| **Villages per arm** | **Quasi-binomial** | **Beta** | **IPTW** | **Naive** |
| 110 | 0.123 (0.116, 0.129) | 0.026 (0.023, 0.030) | 0.009 (0.007, 0.011) | 0.054 (0.049, 0.058) |
| 126 | 0.126 (0.120, 0.133) | 0.026 (0.023, 0.029) | 0.007 (0.005, 0.008) | 0.060 (0.055, 0.065) |

## 3. Supplementary Figures

**Supplementary Figure S1. Power of quasi-binomial regression at different sample sizes and relative reductions in Penta0 rate under different Penta0 rates in the control arm (columns), coefficient sets (rows) and an ICC of 0.22**

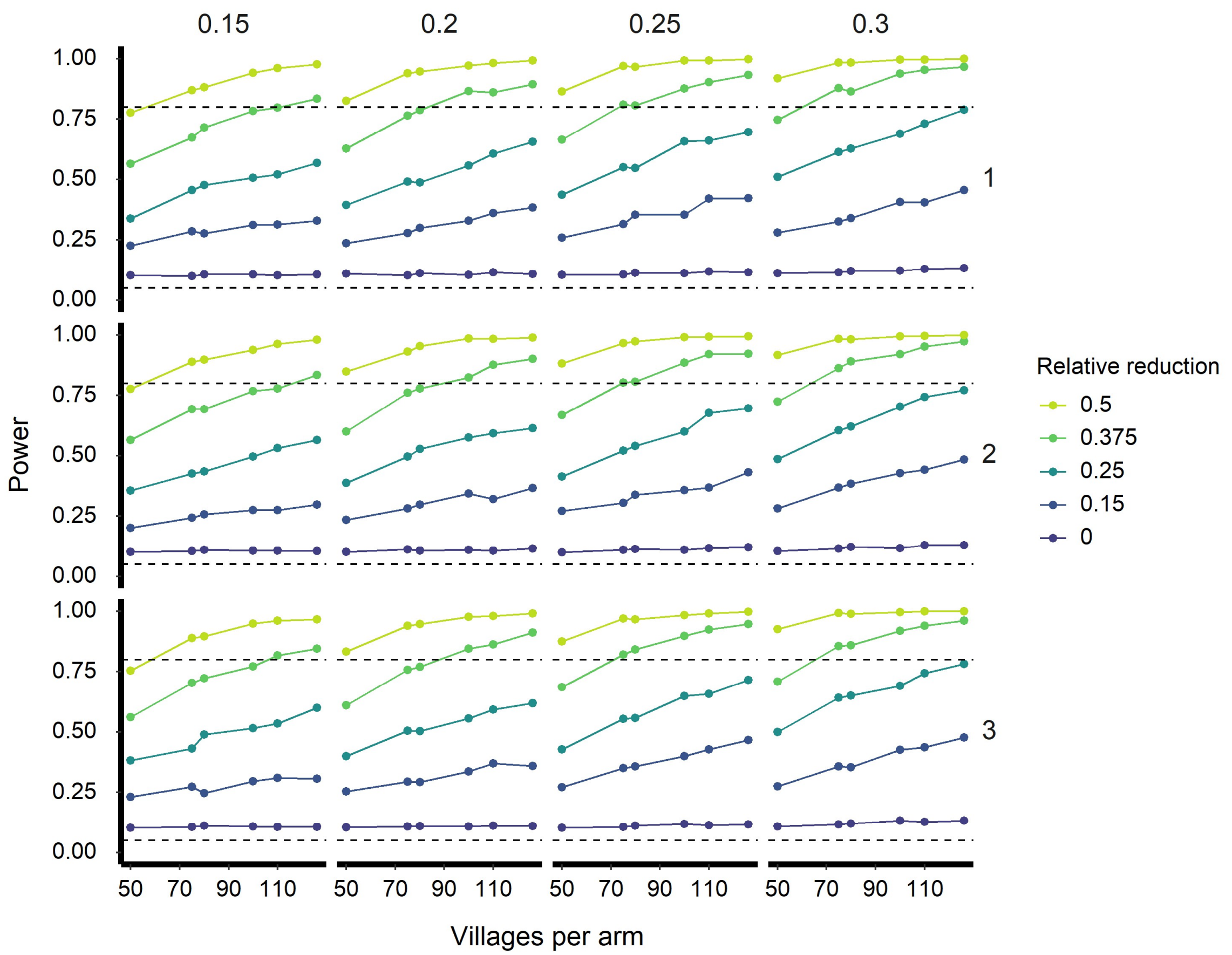

**Supplementary Figure S2. Power of quasi-binomial regression at different sample sizes and relative reductions in Penta0 rate under different Penta0 rates in the control arm (columns), coefficient sets (rows) and an ICC of 1/3**

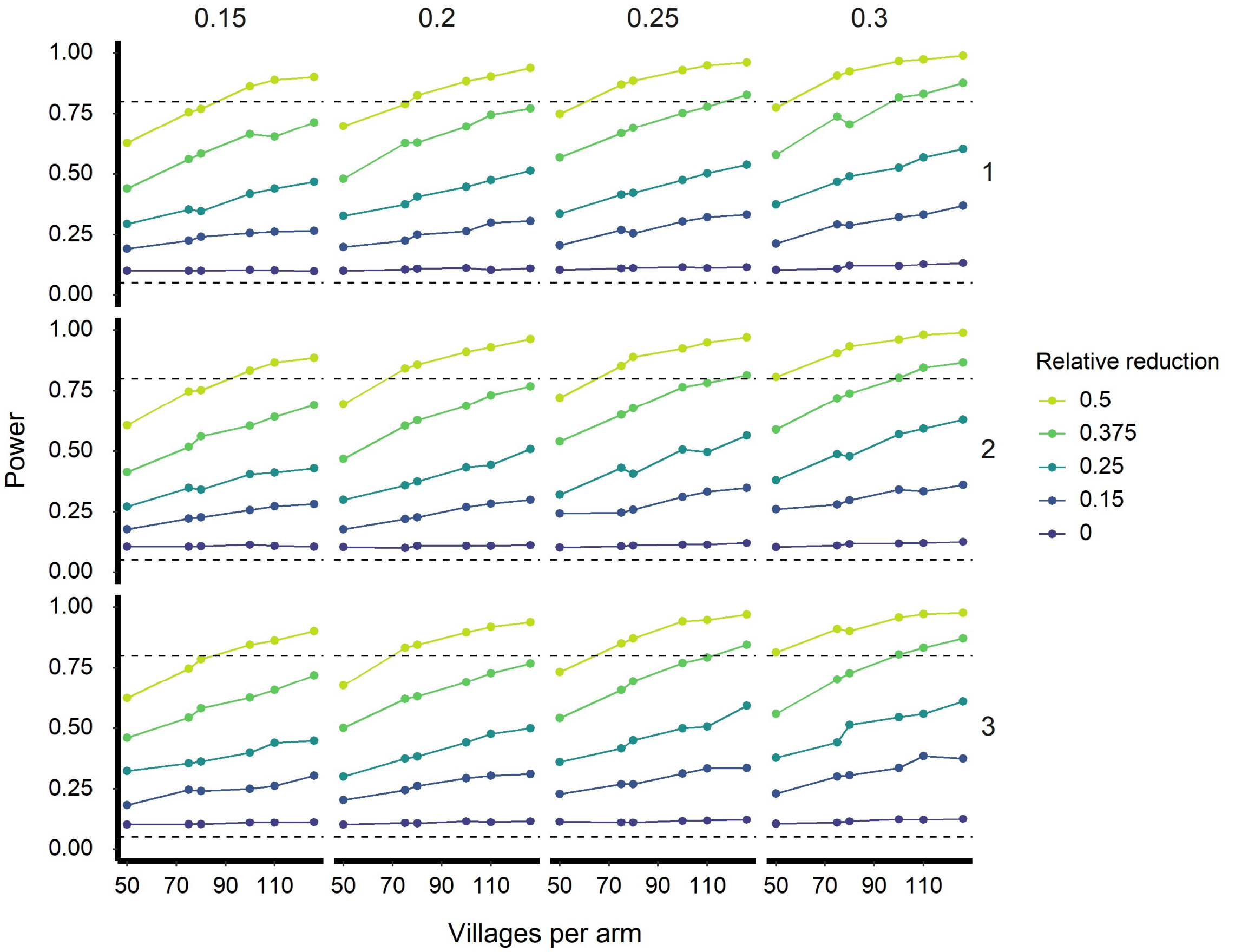

**Supplementary Figure S3. Power of beta regression at different sample sizes and relative reductions in Penta0 rate under different Penta0 rates in the control arm (columns), coefficient sets (rows) and an ICC of 0.22**

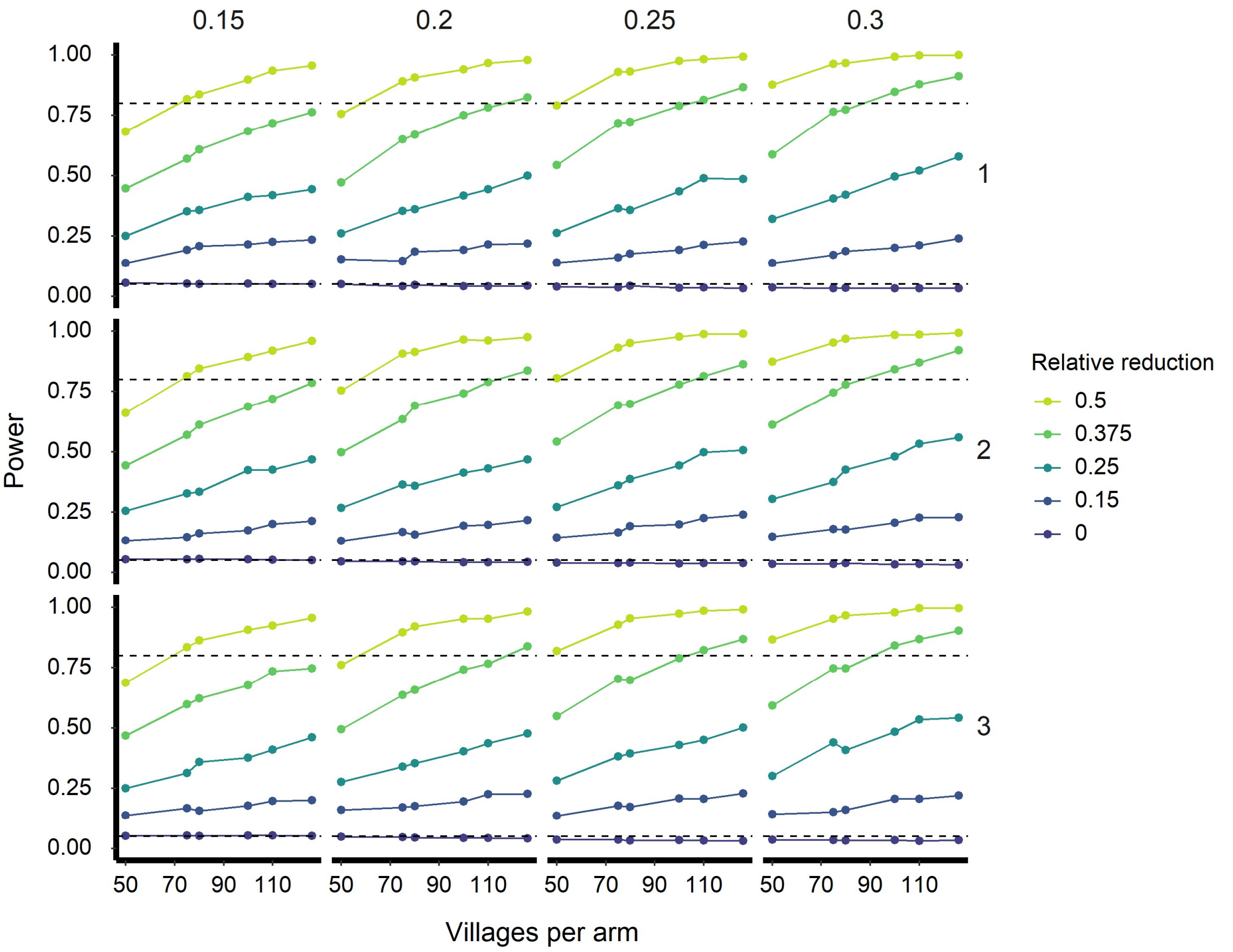

**Supplementary Figure S4. Power of beta regression at different sample sizes and relative reductions in Penta0 rate under different Penta0 rates in the control arm (columns), coefficient sets (rows) and an ICC of 1/3**

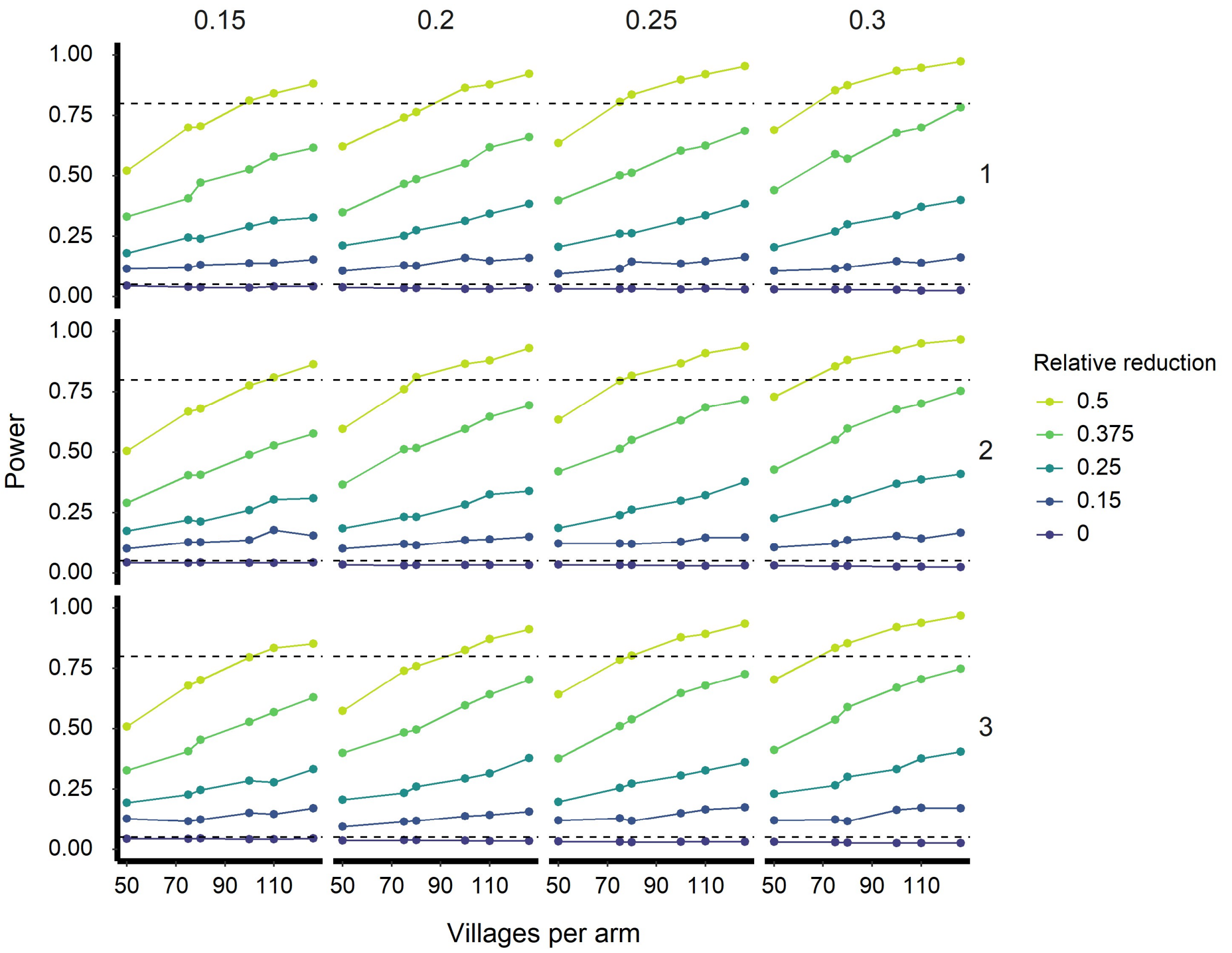

**Supplementary Figure S5. Power of inverse probability of treatment weighting at different sample sizes and relative reductions in Penta0 rate under different Penta0 rates in the control arm (columns), coefficient sets (rows) and an ICC of 0.22**

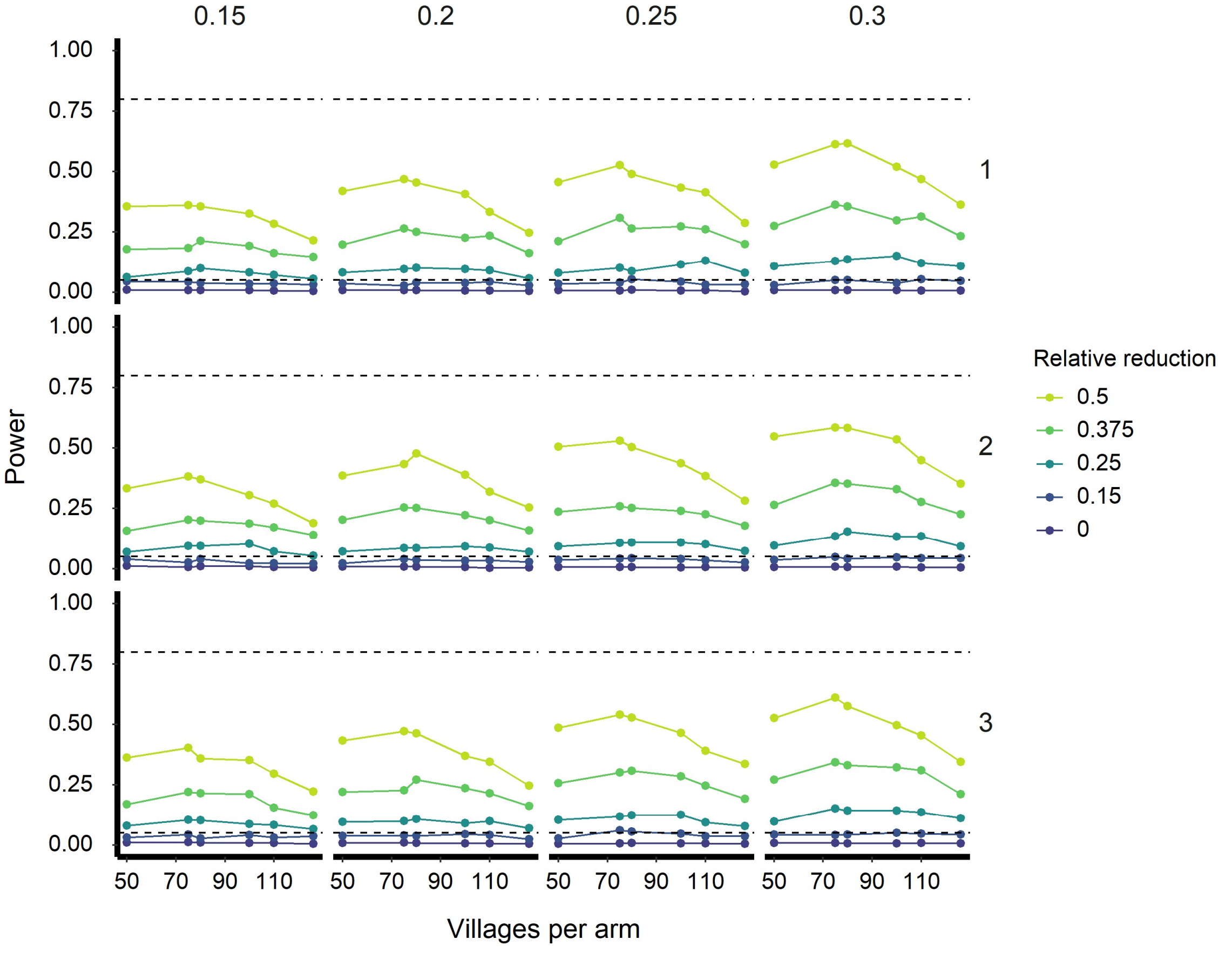

**Supplementary Figure S6. Power of inverse probability of treatment weighting at different sample sizes and relative reductions in Penta0 rate under different Penta0 rates in the control arm (columns), coefficient sets (rows) and an ICC of 1/3**

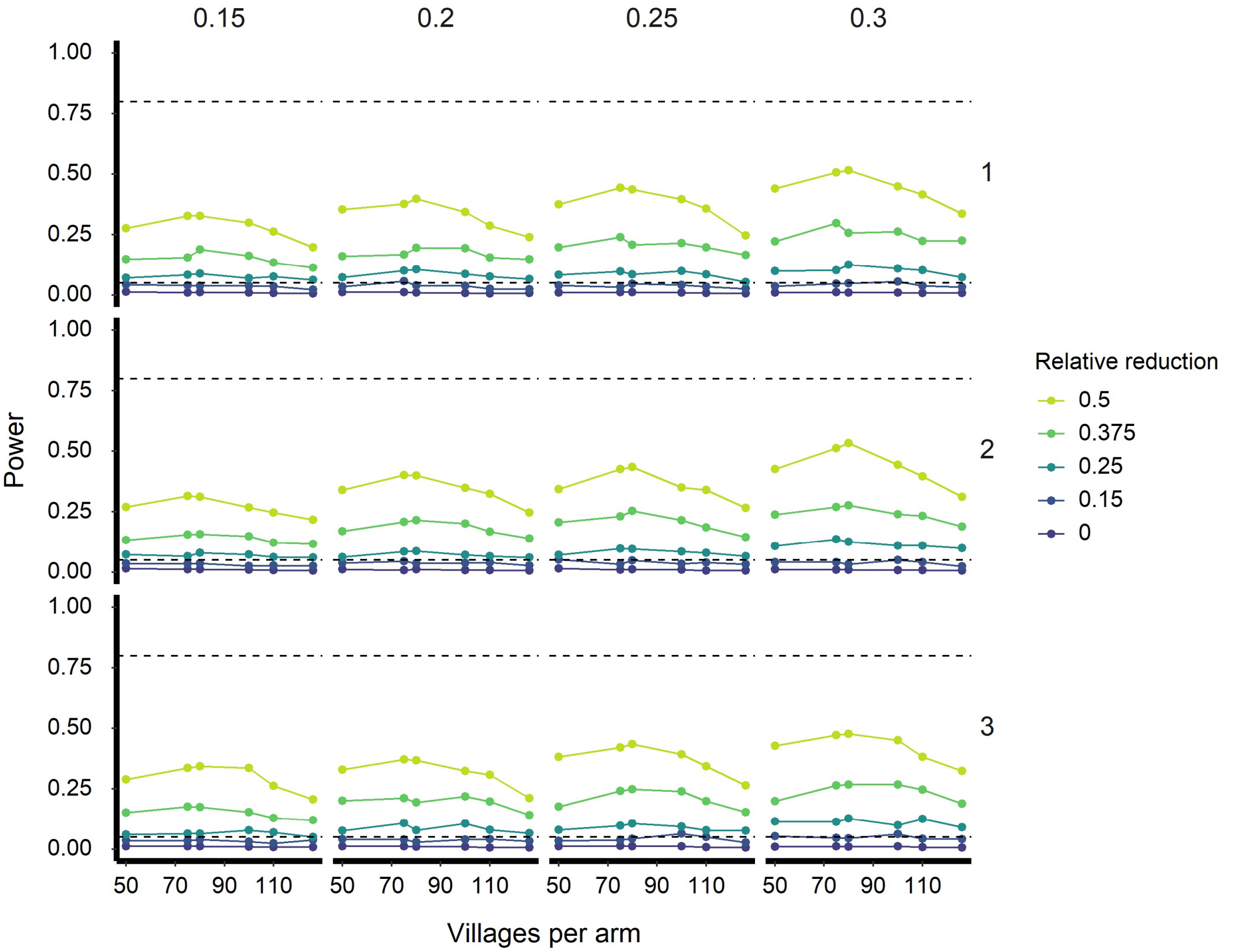

**Supplementary Figure S7. Power of naive analysis at different sample sizes and relative reductions in Penta0 rate under different Penta0 rates in the control arm (columns), coefficient sets (rows) and an ICC of 0.22**

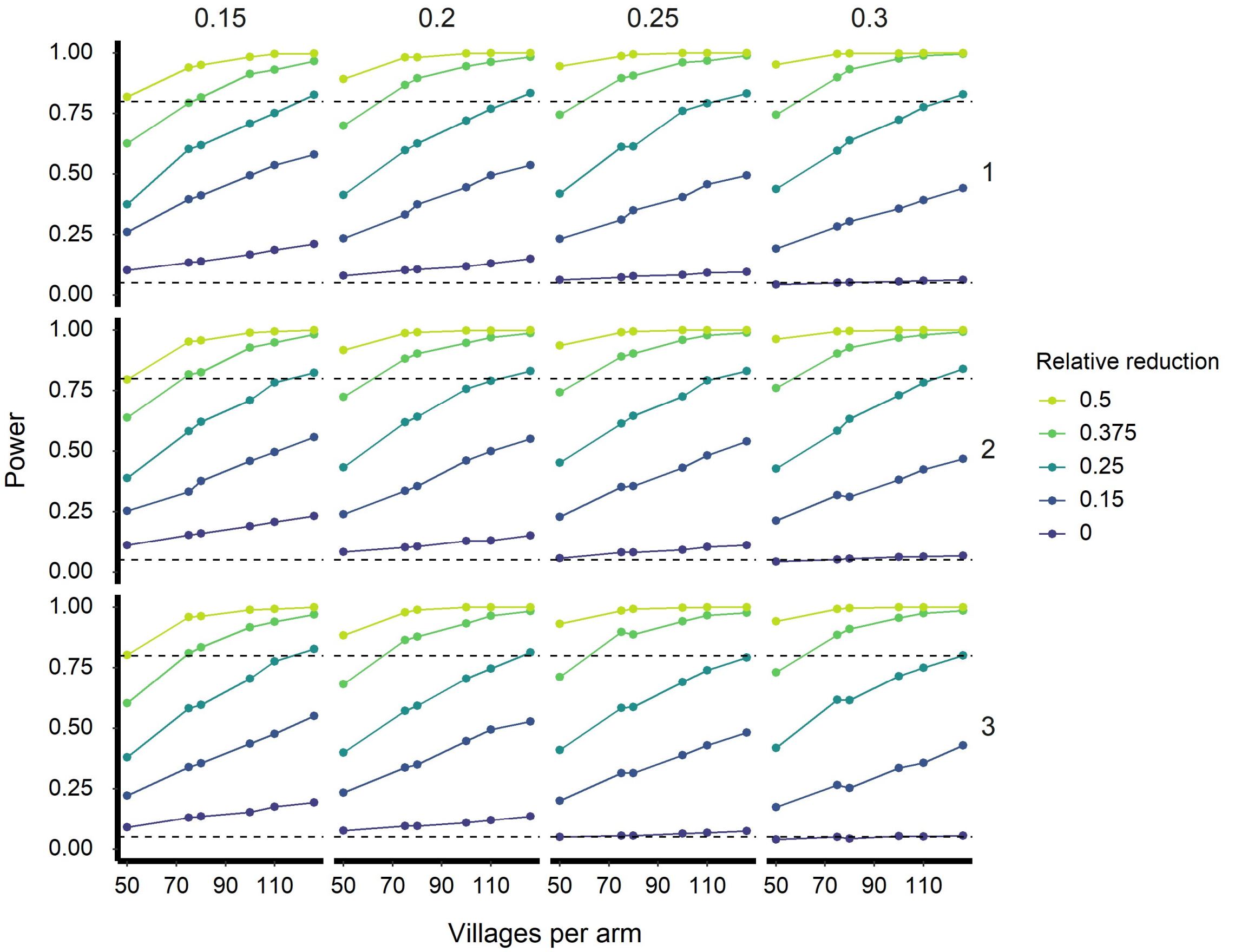

**Supplementary Figure S8. Power of naive analysis at different sample sizes and relative reductions in Penta0 rate under different Penta0 rates in the control arm (columns), coefficient sets (rows) and an ICC of 1/3**

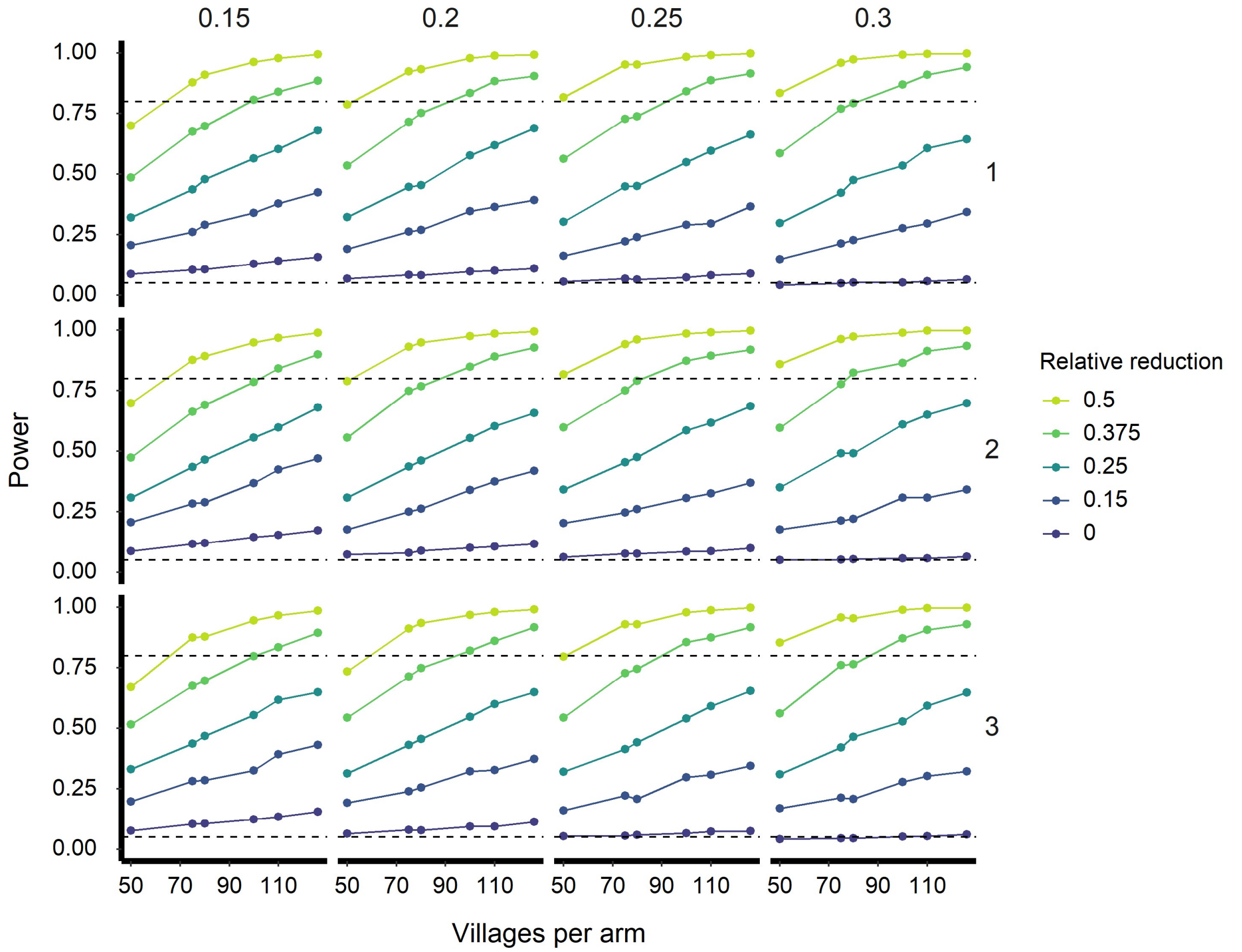